\documentclass[twocolumn,twocolappendix]{aastex701}

\usepackage{amsmath}
\usepackage{enumitem}
\usepackage{comment}

\begin{document}

\title{Dynamical Test of Cosmic Acceleration: $k$-nearest Neighbor Cross Correlation of Cosmic Microwave Background and Cosmic Infrared Background} 

\author[0000-0003-4127-6110]{Dongkok Kim}
\affiliation{Astronomy Program, Department of Physics and Astronomy, Seoul National University, 1 Gwanak-ro, Gwanak-gu, Seoul 08826, Republic of Korea}
\affiliation{Institute for Data Innovation in Science, Seoul National University, Seoul 08826, Korea}
\email[show]{cruithne33@snu.ac.kr}  

\author[0000-0002-8434-979X]{Donghui Jeong}
\affiliation{Department of Astronomy and Astrophysics and Institute for Gravitation and the Cosmos, The Pennsylvania State University, University Park, PA 16802, USA}
\affiliation{School of Physics, Korea Institute for Advanced Study, 85 Heogiro, Dongdaemun-gu, Seoul, 02455, Republic of Korea}
\email{djeong@psu.edu}

\author[0000-0001-6320-261X]{Yi-Kuan Chiang}
\affiliation{Academia Sinica Institute of Astronomy and Astrophysics (ASIAA), No. 1, Section 4, Roosevelt Road, Taipei 10617, Taiwan}
\email{ykchiang@asiaa.sinica.edu.tw}

\author[0000-0003-3428-7612]{Ho Seong Hwang}
\affiliation{Astronomy Program, Department of Physics and Astronomy, Seoul National University, 1 Gwanak-ro, Gwanak-gu, Seoul 08826, Republic of Korea}
\affiliation{SNU Astronomy Research Center, Seoul National University, 1 Gwanak-ro, Gwanak-gu, Seoul 08826, Republic of Korea}
\affiliation{Institute for Data Innovation in Science, Seoul National University, Seoul 08826, Korea}
\email[show]{hhwang@astro.snu.ac.kr}
\begin{abstract}
The physical origin of cosmic acceleration remains one of the central questions in modern cosmology. The integrated Sachs--Wolfe (ISW) and Rees--Sciama (RS) effects, which arise from the evolution of gravitational potentials, provide a dynamical test of cosmic acceleration. We measure the cross correlation between the Planck temperature map of the cosmic microwave background (CMB) and a foreground-free map of the $100\,\micron$ cosmic infrared background (CIB) from \citet{Chiang2023}, using the $k$-nearest neighbor cumulative distribution function ($k$NN-CDF). As the CIB map is reconstructed from ${\sim}600$ million Wide-field Infrared Survey Explorer galaxies extending to $z\approx2.5$, the measurement is equivalent to a galaxy--CMB cross correlation. We find evidence for a positive cross correlation, rejecting the null hypothesis of no correlation at $p=0.02~(2.3\sigma)$ using the $\chi^2$ test. 
We further quantify its amplitude with $A_{k\mathrm{NN}}$, defined relative to an assumed cosmological model, and obtain $A_{k\mathrm{NN}}=0.95\pm0.20~(4.8\sigma)$ with respect to $\Lambda\mathrm{CDM}$ mock data including the ISW and RS effects. The $k$NN-CDF analysis improves the detection significance by ${\sim}20\%$ over the two-point correlation function. We also explore evolving dark energy by constructing two tomographic maps from the full CIB map; the results are consistent with $\Lambda$CDM. Finer tomographic reconstruction of the CIB, for example from multiband data of SPHEREx, would further tighten constraints on evolving dark energy.

\end{abstract}

\keywords{\uat{Cosmology}{343} --- \uat{Cosmic background radiation}{317} --- \uat{Dark energy}{351}}


\section{Introduction \label{sec:Sec1}}
The cosmic microwave background (CMB) has been central to modern cosmology ever since its discovery in the 1960s \citep{PW1965}. The minute anisotropies in their energy distribution enabled precise measurements of the initial conditions of the Universe through successive experiments \citep{Bennett2003, Planck2011}, establishing the foundation of the $\Lambda$CDM model. Beyond the primordial fluctuations released at $z\sim1100$, the interactions of CMB photons with large-scale structures at later times open an additional avenue to test the cosmological model and to probe dark energy.

Dark energy, responsible for the accelerated expansion of the late-time Universe, constitutes roughly $70\%$ of the present-day cosmic energy density. Its existence and basic properties have been constrained by a variety of measurements, including Type Ia supernova luminosity distances \citep{Riess1998, Perlmutter1999}, large-scale structure surveys \citep{eBOSS2021, DES2022, Adame2025}, and CMB observations \citep{Komatsu2011, Planck2020}. These probes predominantly test the geometrical effects of dark energy on the distance–redshift relation and the growth of structure. Yet, despite this extensive evidence, the physical nature of dark energy remains unknown, and even fundamental quantities such as its equation of state $w$ remain under active debate \citep[e.g.,][]{Dong2023, Adame2025}. 

This motivates the need for complementary observables—particularly those that provide a dynamical test of dark energy. Such probes add independent information by directly tracing the growth of the large-scale structure. While geometrical measurements primarily constrain integrals of the Hubble parameter $H(z)$ through the luminosity distance or the angular-diameter distance, dynamical measurements directly access the effect of Hubble damping on the growth of large-scale structure through the peculiar velocity or the evolution of the gravitational potential.

One such dynamical probe is the integrated Sachs--Wolfe (ISW; \citealt{SachsWolfe1967}) effect and its nonlinear extension, the Rees--Sciama (RS; \citealt{RS1968}) effect. Hereafter, we use ``ISWRS'' to denote the combined secondary anisotropy including both the ISW (linear) and RS (nonlinear) effects. These effects arise from the gravitational redshift experienced by CMB photons as they propagate through gravitational potentials that evolve with cosmic time. CMB photons crossing a static potential gain no net energy because the gravitational redshift and blueshift cancel, whereas a time-varying potential yields a net imprint on the observed CMB temperature. In the matter-dominated era the linear gravitational potential stays constant, but it evolves once dark energy becomes dynamically important, producing CMB anisotropies correlated with large-scale structure.

The typical temperature anisotropies generated by the ISWRS effect are an order of magnitude smaller than the primary CMB fluctuations. Consequently, specialized detection strategies, most notably (1) cross correlation with the large-scale density field and (2) stacking analyses of nonlinear structures such as clusters and voids, have been developed and applied to wide-field surveys. 
For example, cross-correlation analyses between the CMB and tracers of large-scale structure have been performed using the Sloan Digital Sky Survey \citep[SDSS;][]{HM2014}, the Two Micron All Sky Survey (2MASS), the NRAO VLA Sky Survey (NVSS), the X-ray background \citep[see][and references therein]{Dupe2011}, the Wide-field Infrared Survey Explorer \citep[WISE;][]{Goto2012, Ferraro2015, Krolewski2022}, and the DESI Legacy Survey \citep{Dong2021, Hang2021a}, as well as through combinations of multiple surveys \citep{Ho2008, G2012, Stolzner2018}. Note that the ISWRS signal is subtle, and previous studies have achieved a significance of $\gtrsim 4\sigma$ only when multiple surveys are jointly analyzed.
For the stacking analysis, \citet{Granett2008}, \citet{Nadathur2012}, \citet{Flender2013}, \citet{Ilic2013}, \citet{HM2013}, \citet{Cai2014}, \citet{Planck2014}, \citet{Hotchkiss2015}, \citet{Nadathur2016}, \citet{Planck2016}, \citet{Hang2021b}, and \citet{Kovacs2022} have stacked CMB images at the locations of clusters/voids to detect hot/cold spots; these studies yield $\lesssim 3\sigma$ detection. 

Alongside these observational efforts, substantial work has been devoted to refining theoretical predictions and analysis techniques.
Linear theory calculations \citep{Afshordi2004} and fully nonlinear $N$-body simulations \citep{Cai2009, Cai2010, Hotchkiss2015} have been used to model the ISWRS contribution, identify where it dominates over noise and primary anisotropies, and quantify how survey depth, sky fraction, and tracer bias affect detectability. Building on these modeling efforts, several works have proposed optimal filtering and combination techniques \citep{Cabre2007, Frommert2008, Ballardini2019, Ferraro2022} that maximize the signal-to-noise ratio by appropriately weighting multipoles or density tracers.

In this work, we measure the ISWRS signal using the cosmic infrared background (CIB) as a tracer of large-scale structure. The CIB is the integrated infrared emission from dusty star-forming galaxies and traces large-scale structure over a broad redshift range, so is well suited for ISW detection (see \citealp{Maniyar2019} for forecasts; \citealp{Chiang2025} for the redshift distribution). Constructing a reliable CIB map over wide areas has been difficult due to strong foreground contamination from Galactic dust. We overcome this by using the foreground-cleaned $100\,\micron$ CIB reconstruction of \citet[][hereafter C23]{Chiang2023}, which isolates the extragalactic component from the \citet[][hereafter SFD]{SFD1998} Galactic dust map (based on the IRAS intensity map) through cross correlations with spectroscopic sources. The resulting map, built from density fields of ${\sim}600$ million WISE galaxies extending to $z\approx2.5$, provides a near full-sky tracer of large-scale structure for ISWRS analyses.

To achieve a detection with a high signal-to-noise ratio (S/N), we adopt a new cross-correlation method based on the $k$-nearest neighbor cumulative distribution function ($k$NN-CDF). 
In brief, the $k$NN statistic measures the empirical cumulative distribution of distances from random points to their $k$th nearest neighbor in the discrete dataset. The $k$NN-CDF is sensitive to an integrated combination of higher-order clustering information, which can capture the clustering signal more efficiently than classical two-point estimators. This motivates its application to the ISWRS problem, where higher-order clustering information can help improve the detection significance. We present the formalism and covariance estimation in Section \ref{sec:Sec3}.

Using the Planck CMB temperature map and the C23 $100\,\micron$ CIB map, we measure the ISWRS-induced cross correlation with the $k$NN-CDF and quantify its detection significance relative to $\Lambda$CDM mock data that include the ISWRS effects. We compare the $k$NN-CDF result with a two-point analysis to assess potential gains in significance. To probe the possibility of evolving dark energy, we further split the full CIB map into two components that trace different redshifts, and we tomographically cross-correlate them with the CMB map.

This paper is organized as follows. In Section \ref{sec:Sec2}, we describe the datasets used in this study and the mock data constructed from an $N$-body simulation. Section \ref{sec:Sec3} details the $k$NN measurement framework and covariance estimation. We present the detection results in Section \ref{sec:Sec4}. We discuss future prospects and tomographic information in Section \ref{sec:Sec5}. We conclude in Section \ref{sec:Sec6}. \\ 

\section{Data \label{sec:Sec2}}
In this section, we describe the data we use for the analysis. The ISWRS measurement requires low-redshift galaxy data and CMB data. To compare the measurement with the $\Lambda$CDM prediction, we generate the mock CIB map and ISWRS map using the simulation data.

\subsection{Observational Data \label{sec:Sec2.1}}
\subsubsection{$100\,\micron$ CIB map \label{sec:Sec2.1.1}}

\begin{figure*}[t]
\centering
\includegraphics[width=0.99\textwidth]{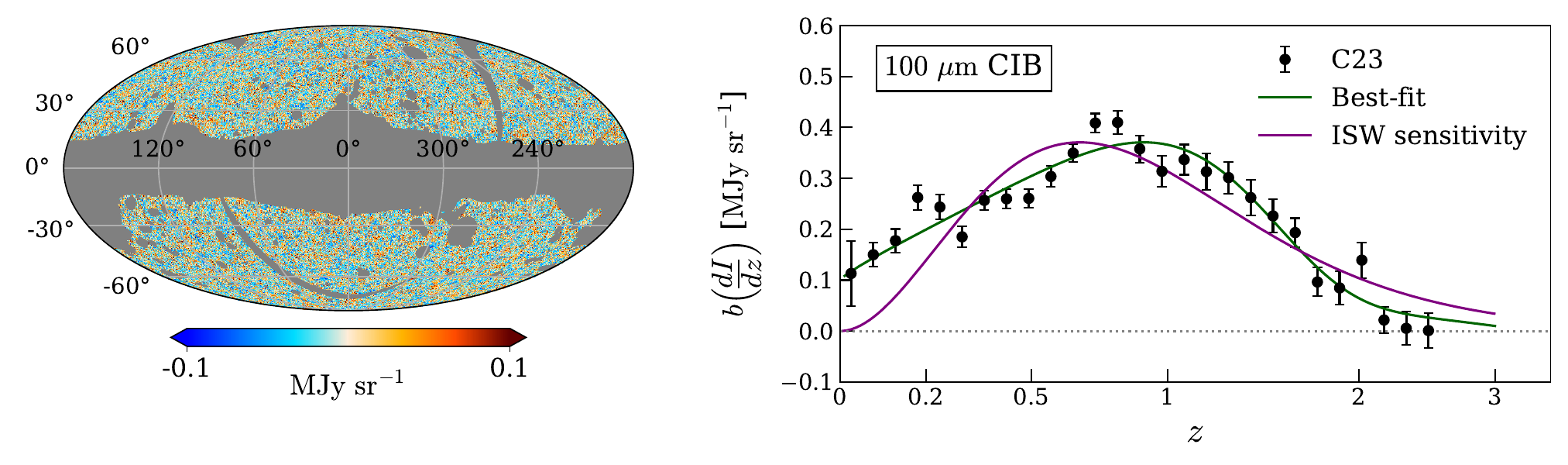}
\caption{Left: $100\,\micron$ CIB extracted from the SFD dust map in Galactic coordinate. Right: bias-weighted redshift distribution of $100\,\micron$ CIB. The black data points denote the measurements using the clustering redshift method of \citet{Chiang2025}. The green line shows the smoothed distribution using the analytical function described in Equation \ref{eq:Eq1}. The purple line shows the ISW sensitivity function defined as $\propto \frac{d[(1+z)D(z)]}{dz} \frac{dV}{dz}$.
\label{fig:Fig1}}
\end{figure*}

As the matter tracer of the late Universe, we use the reconstructed $100\,\micron$ CIB anisotropy map from C23 (Figure \ref{fig:Fig1}; left panel), derived from galaxy density fields traced by ${\sim}600$ million WISE sources over most of the extragalactic sky. This map isolates the CIB component superimposed on the bright Milky Way dust emission in the SFD $100\,\micron$ map, thereby providing a clean tracer of large-scale structure for cosmological analyses.

To perform this component separation, C23 measured tomographic angular cross correlations between SFD and spectroscopic reference galaxy and quasar samples from SDSS \citep{York2000}, in a manner similar to the clustering redshift approach \citep{Menard2013}, thereby extracting the redshift- and scale-dependent two-point statistics. Constrained by this information, the CIB map is reconstructed as a linear combination of galaxy density templates drawn from three catalogs: (1) 18 samples from CatWISE2020 \citep{Marocco2021} selected with different color and magnitude cuts, (2) 12 samples from the WISE$\times$SuperCOSMOS catalog \citep{Bilicki2016} divided by photometric redshift, and (3) a sample of AKARI point sources \citep{Yamamura2018}, detected at similar wavelengths and representing part of the Poisson contribution to the CIB just below the IRAS point-source detection limit. To better capture small-scale structure, C23 augments the first two sets with six beam variants each, yielding $(180+1)$ templates in total. The linear coefficients are chosen to reproduce the full set of tomographic two-point statistics of the CIB measured from SFD. In this framework, the templates encode the phase information of the large-scale structure, while the cross-correlation measurements define the target power spectrum and redshift distribution. The reconstruction is maximally empirical, requiring no explicit modeling of the properties of CIB sources, matter clustering, or the galaxy bias of either the CIB or the reference samples.

The CIB map is provided in HEALPix format with $N_{\mathrm{side}}=2048$ \citep{Gorski2005}, which sufficiently samples the SFD resolution\footnote{Available together with CSFD, the CIB-cleaned Milky Way dust map is at DOI:\dataset[10.5281/zenodo.8207175]{https://doi.org/10.5281/zenodo.8207175}}. It contains only anisotropy, with no monopole by construction. In this work, we use a tailored version that excludes the Poisson contribution from AKARI point sources. As these sources are predominantly at low redshift ($z<0.1$), their removal reduces noise while leaving the CIB--ISW cross-correlation signal largely unaffected. We estimate the covariance and associated uncertainties using 200 bootstrap realizations.

For subsequent modeling, we estimate the bias-weighted redshift distribution of the CIB, $b\,dI/dz$, using the clustering redshift formalism. Specifically, we divide the measured cross-correlation function---integrated over the adopted angular scales to maximize the clustering signal---by the matter autocorrelation function and the bias of the reference sample. Details of the adopted angular scales and weighting scheme for the cross correlation are provided by C23 and \citet{Chiang2025}.
The right panel of Figure \ref{fig:Fig1} shows the bias-weighted redshift distribution of the $100\,\micron$ photons together with the ISW sensitivity function defined as $\propto \frac{d[(1+z)D(z)]}{dz} \frac{dV}{dz}$, where $D(z)$ is the linear growth factor and $V$ is the comoving volume. The $100\,\micron$ CIB is particularly well suited for detection of the ISW effect as the bias-weighted redshift distribution peaks at $z\approx1$ and extends up to $z\approx2.5$. To obtain a smooth redshift distribution, we use an empirical fitting formula composed of three parameters. The function is as follows:
\begin{equation}
    \label{eq:Eq1}
    b\frac{dI}{dz} \propto z^{\alpha} \exp\left[ -\left( \frac{z}{z_0} \right)^{\beta} \right] + az + b
\end{equation} 
where the first term is similar to the empirical fitting function widely used for a flux-limited galaxy survey and the linear term with $(a,b)=(-0.03,0.1)$ is arbitrarily adopted to better describe the low-redshift emission ($z\lesssim0.2$). \\

\subsubsection{CMB Temperature Map \label{sec:Sec2.1.2}}
For the CMB data, we use the temperature anisotropy map from the Planck Public Data Release 3 \citep[PR3;][]{Planck2020-1}. In particular, we adopt the product of the SMICA component-separation pipeline \citep{Planck2020-2}, which additionally provides a ``Sunyaev--Zel'dovich (SZ)-free'' CMB temperature map (SMICA-nosz). The SZ effect is generated by hot gas residing in deep gravitational potentials, typically regions where galaxies cluster. Although the SZ effect is spectrally distinct from the ISWRS effect, it is known to leave cold-spot residuals in Planck CMB maps, potentially impacting the CMB-density cross correlation on small angular scales \citep{Bobin2016, Chen2018}. By using the ``SZ-free'' temperature map, we reduce the influence of SZ contamination on our cross-correlation analysis. \\

\subsubsection{Mask \label{sec:Sec2.1.3}} 
The CMB and CIB maps are contaminated by foreground emission. Along most lines of sight---especially at low Galactic latitudes---these foregrounds are orders of magnitude stronger than the cosmological signals. To mitigate this contamination, we apply a combined mask defined as the union of the CIB mask from C23 and the Planck common mask \citep{Planck2020-2}. The resulting mask is shown in gray in the left panel of Figure \ref{fig:Fig1}. The CIB mask was constructed by merging the mask tied to the galaxy templates described in Section \ref{sec:Sec2.1.1} with a stellar-density mask derived from Gaia DR3 \citep{Gaia2023}. It was ultimately partitioned into cosmology and noncosmology regions. We restrict our analysis to the cosmology region and additionally remove areas lacking IRAS coverage so as to ensure a consistent effective resolution across the sky. The Planck common mask---based on interpipeline standard deviations and point sources---largely overlaps the CIB mask ($<1\%$ difference) and leaves about $56\%$ of the sky unmasked. 

We further exclude pixels near the mask boundaries. Because the $k$NN statistic uses top-hat smoothed CMB and CIB maps (Section \ref{sec:Sec3}), we dilate the baseline mask by $\theta$, the top-hat radius, to prevent mixing with masked pixels.

\subsection{Simulation \label{sec:Sec2.2}}
To compare the observation with $\Lambda$CDM prediction, we construct mock data from the publicly available Millennium-XXL (MXXL) simulation \citep{Angulo2012}. MXXL is an $N$-body simulation containing $6730^3$ dark matter particles in a $3\,h^{-1}\,\mathrm{Gpc}$ volume conducted with the GADGET-3 code. The simulation adopts cosmological parameters of $H_0=73\,\mathrm{km}\,\mathrm{s}^{-1}\,\mathrm{Mpc}^{-1}$, $\Omega_{m,0}=0.25$, $\Omega_{\Lambda,0}=0.75$, $\sigma_8=0.9$, and $n_s=1$, following the original Millennium simulation \citep{Springel2005}.
The large volume of the simulation enables us to simulate the full sky up to $z\sim2.7$ using a $3^3$ tiling of the simulation box, which covers most of the redshift range of the $100\,\micron$ CIB emission (see Figure \ref{fig:Fig1}). The resolution enables the dark matter halos to be resolved down to ${\sim}10^{11.5}\,M_{\sun}$ with 50 particles, so most of the halos contributing to the $100\,\micron$ CIB are included \citep{PlanckCIB2014}. \\

\subsubsection{Mock CIB map \label{sec:Sec2.2.1}}
\begin{figure*}[t]
\centering
\includegraphics[width=0.99\textwidth]{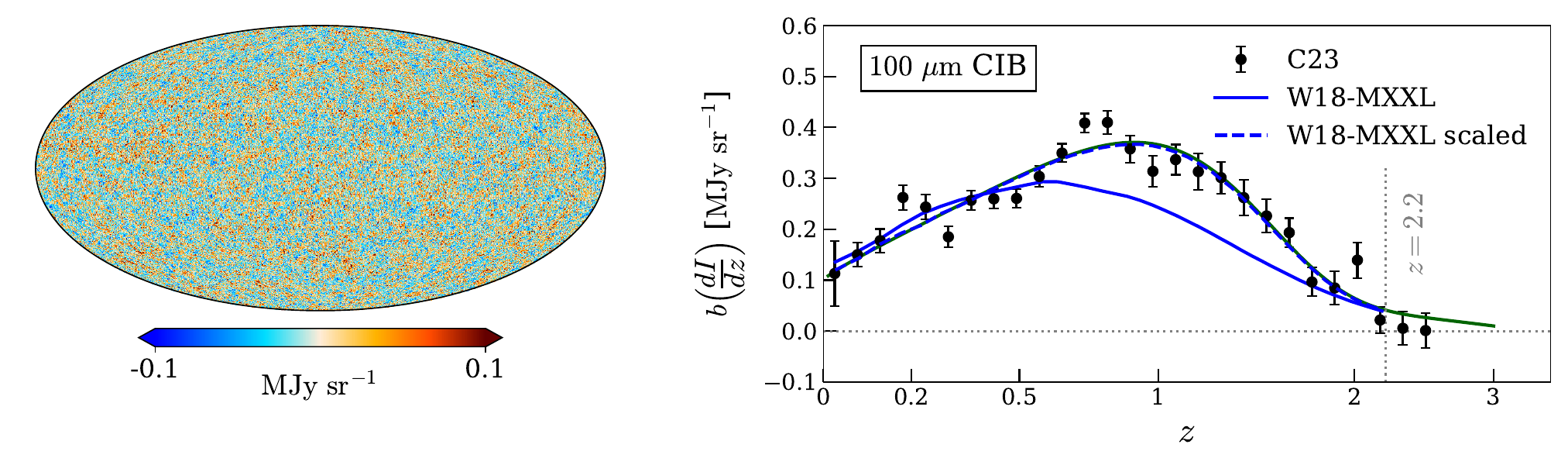}
\caption{Left: mock $100\,\micron$ CIB map constructed with MXXL halos. The map is smoothed using the transfer function, and white noise terms are added (see Section \ref{sec:Sec2.2.1} and Figure \ref{fig:Fig4}). Right: same as Figure \ref{fig:Fig1}, but with a mock redshift distribution calculated using the W18 model (solid and dashed blue lines). The model agrees with the observation within $30\%$ at low redshift ($z<0.8$), but requires a larger scaling factor up to $60\%$ at higher redshift. The boundary of the lightcone halo catalog ($z=2.2$) is also shown as a gray dotted vertical line. The absence of halos at $z>2.2$ is negligible compared to the total radiation.
\label{fig:Fig2}}
\end{figure*}

\begin{figure}[t]
\centering
\vspace{5mm}
\includegraphics[width=0.93\linewidth]{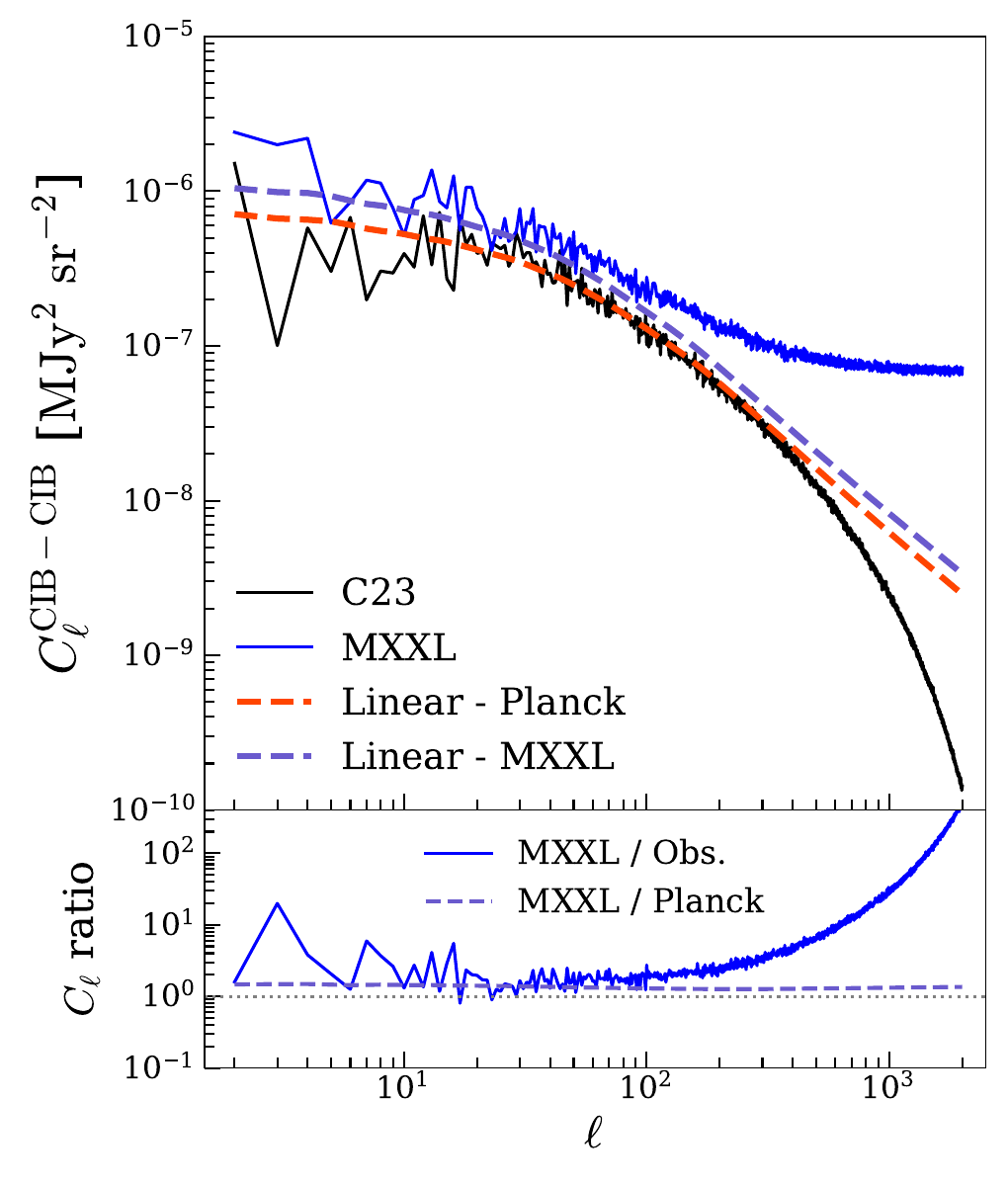}
\caption{Top: angular power spectrum of $100\,\micron$ CIB from C23 (black) and MXXL simulation (blue). The fluctuations are larger on a large scale in MXXL. There is also a shape difference on a small scale due to the absence of a beam function. Power spectra in the linear theory (dashed) are included to demonstrate the impact of assumed cosmology. The power spectrum is ${\sim}10\%\text{--}20\%$ larger with MXXL cosmology. Bottom: ratio between two power spectra. The blue solid line denotes the ratio between MXXL and observation; the dashed line shows the ratio between MXXL and Planck cosmologies. The ratio between simulation and observation on a large scale is similar to the difference in the two cosmologies.  
\label{fig:Fig3}}
\end{figure}

\begin{figure*}[t]
\centering
\includegraphics[width=0.9\textwidth]{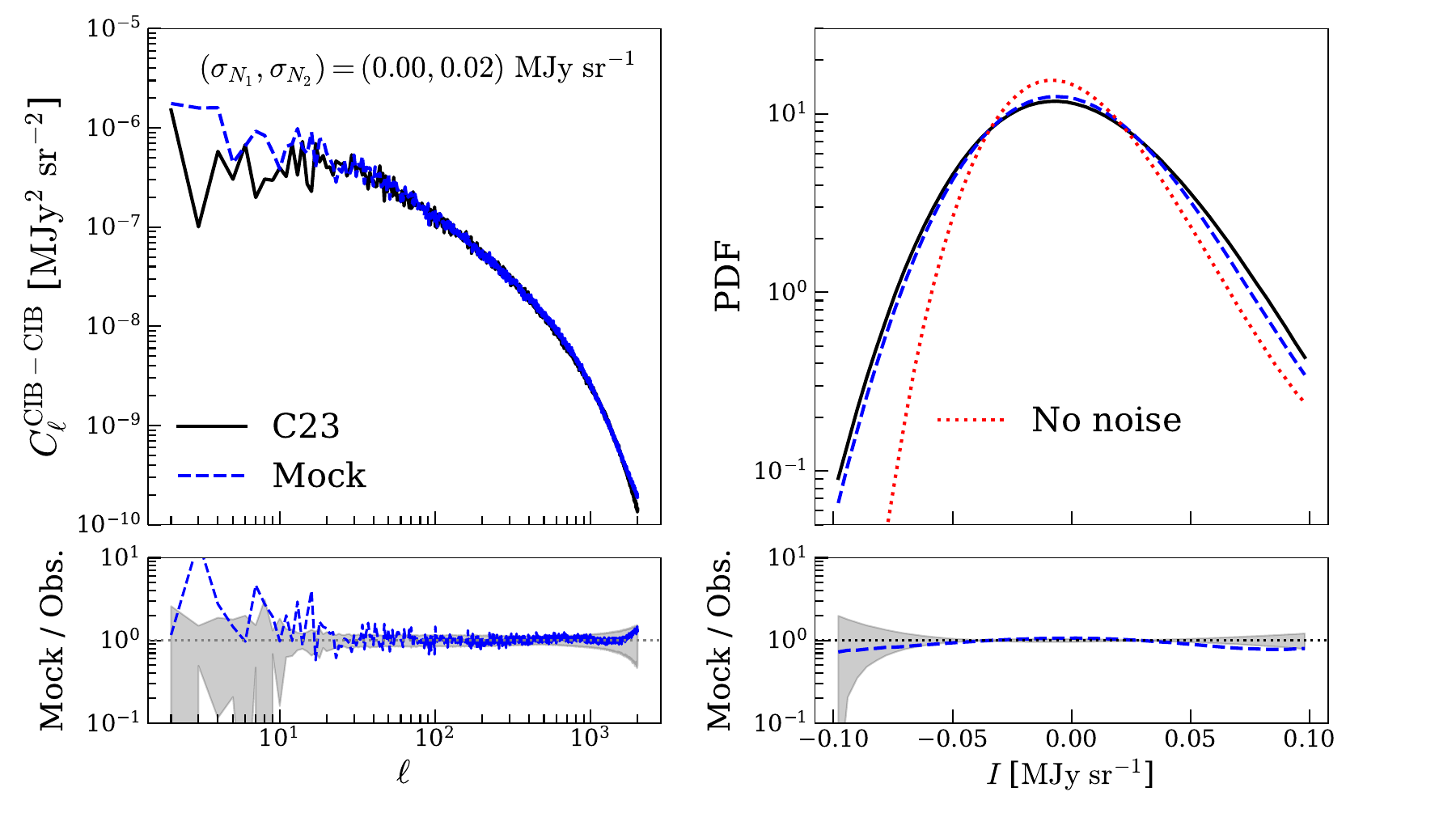}
\caption{Left: angular power spectrum of $100\,\micron$ CIB. Right: PDF of CIB intensity. The black lines denote the observation and blue lines denote the readjusted mock data. To match both one-point and two-point statistics, white noise terms of $(\sigma_{N_1}, \sigma_{N_2})=(0, 0.02)\,\mathrm{MJy}~\mathrm{sr}^{-1}$ are added. The ratios between mock and observational data are shown in the bottom panels. The gray shading shows the range of the standard deviation estimated with 200 realizations of the CIB. The mock and the data agree within $1\sigma$. The mock data without noise terms are shown as a red dotted line to illustrate the impact of white noise.
\label{fig:Fig4}}
\end{figure*}
Our use of the reconstructed CIB map is advantageous in the context of baryon painting. Constructing mock data for the WISE galaxy templates would require explicitly populating dark matter halos with galaxies, for example through a halo occupation model or by assigning luminosities in the relevant passbands. It would further require matching the WISE selection function. In contrast, the CIB requires fewer modeling ingredients, because it can be described by painting dust emission onto the simulated large-scale structure with an existing CIB prescription. Both approaches still require tuning the mocks to match the observed statistics, but the CIB avoids the additional steps of populating halos with individual galaxies and matching a galaxy selection function.

To paint the CIB on MXXL halos, we adopt the halo infrared emission model by \citet[][hereafter W18]{Wu2018}. The model uses a gas regulator framework to parameterize the baryon cycle, star formation activity, and hence infrared luminosity of a galaxy. Infrared luminosities are assigned to halos as a function of mass and redshift.
W18 calculates the specific intensity assuming a modified blackbody (MBB) spectral energy distribution for dust emission; this assumption, however, breaks down at higher frequencies such as $100\,\micron$ due to warm dust and heterogeneous emission mechanisms \citep{Casey2012}. Since we focus on a single frequency band (3 THz), we bypass the SED modeling and simply scale $b\,dI/dz$ to match the observation.

To construct a mock CIB map, we apply the model to the publicly available MXXL lightcone halo catalog \citep{Smith2017}. The lightcone halos are interpolated between adjacent snapshots following the prescription of \citet{Merson2013} to determine the positions and velocities at the observer lightcone. The observer is located at $(x,y,z)=(17,163,2771)\,h^{-1}\,\mathrm{Mpc}$, and the halo catalog is available up to $z=2.2$. To scale the (W18 model+MXXL) to match the observed $b\,dI/dz$ (Figure \ref{fig:Fig1}), we employ the mass function and bias measured directly from the lightcone halo data.
We measure the halo mass function $dn(M,z) /d\ln M$ and linear bias $b(M,z)$ at the 21 redshifts in the range of $z=[0.1,2.1]$ using a spherical shell of width $\Delta z=0.1$ around each redshift. Details of the measurement procedure and results are given in Appendix \ref{sec:AppA}.
In Figure \ref{fig:Fig2}, we present the model redshift distribution from MXXL that is based on the W18 model. The model (blue solid line) is consistent with the observation at low redshift ($z<0.8$) within $30\%$. However, the model deviates from the observation by as much as ${\sim}60\%$ at higher redshift as the rest-frame wavelength decreases. As noted above, we directly scale the model to match the observation (blue dashed line). Although the CIB extends slightly beyond $z=2.2$, where our lightcone halos end, the contribution from this redshift range is negligible compared to the total radiation.

\begin{figure*}[t]
\centering
\includegraphics[width=0.99\textwidth]{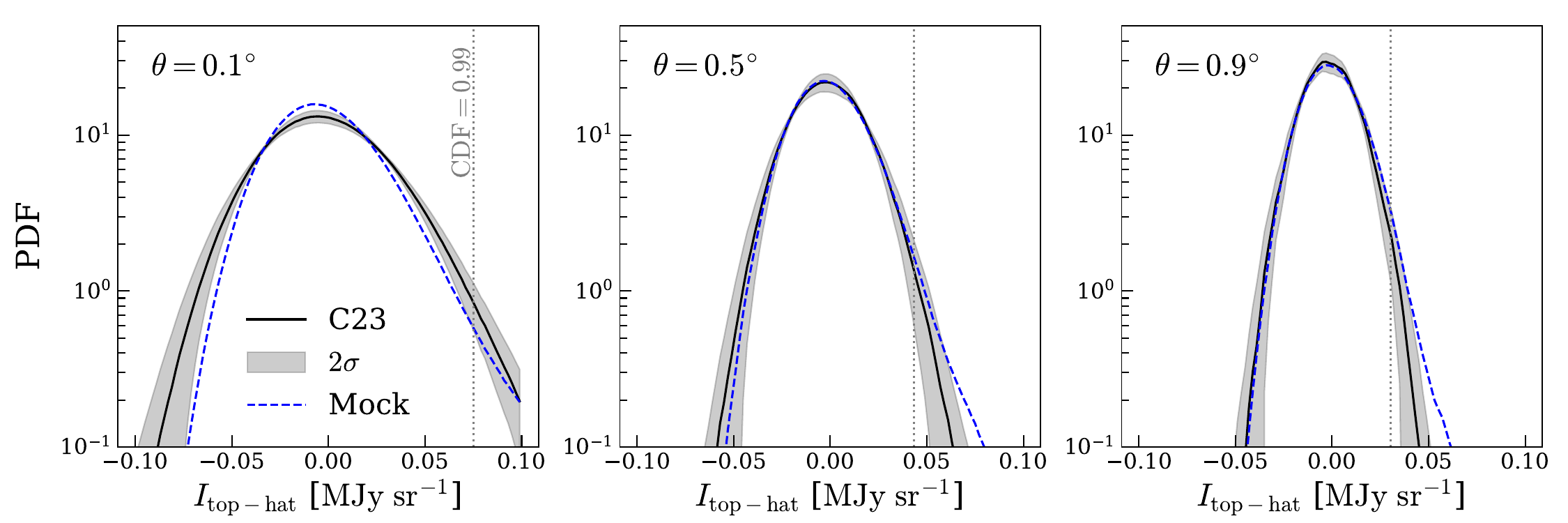}
\caption{PDF of top-hat smoothed CIB intensity. Smoothing scales of $\theta=0.1,0.5,0.9^\circ$ are shown from left to right. The black lines denote the measurements from C23 and the shading shows the $2\sigma$ range calculated with 200 realizations. The results from the mock data are shown as blue dashed lines. C23 and the mock data agree well within uncertainty, particularly at $\mathrm{CDF}<0.99$ (gray dotted vertical line). 
\label{fig:Fig5}}
\end{figure*}

Although the redshift distribution, or angular average of the map, is scaled to match the observation, there remain several discrepancies at the map level. First, the cosmology assumed in the MXXL simulation differs from that of more recent observational constraints, such as those from Planck 2018 \citep[][hereafter Planck cosmology]{Planck2020}. Even assuming the same $b\,dI/dz$, the angular power spectrum of the CIB varies up to ${\sim}10\%\text{--}20\%$ between two cosmologies. The second is the small-scale difference arising from the beam of SFD and imperfect modeling of halos on a small scale (e.g., resolution of the simulation, interpolation error in lightcone space). Lastly, there is additional noise in observation such as shot noise and instrumental noise. In Figure \ref{fig:Fig3}, we show the CIB auto power spectrum from C23 and MXXL along with the linear theory calculated assuming Planck and MXXL cosmologies. The power spectra in the linear theory are calculated using the Limber approximation\footnote{We note that the Limber approximation adopted throughout is accurate to within 1\% on the angular scales used in our analysis.} \citep{Limber1953, LoVerde2008} as follows:
\begin{equation}
    \label{eq:Eq2}
    C_{\ell}^{\mathrm{CIB-CIB}} = \frac{1}{c^2} \int \frac{dr}{r^2} H^2 \left( b\frac{dI}{dz} \right)^2 P \left( k=\frac{\ell+1/2}{r},z \right)
\end{equation}
where $c$ is the speed of light, $r$ is the comoving distance, and $P(k,z)$ is the matter power spectrum at $z$. We use the HALOFIT \citep{Takahashi2012} nonlinear matter power spectrum from the CAMB library \citep{Lewis2000} as our $P(k,z)$. There are two noticeable differences between C23 and MXXL. First, the fluctuations in MXXL are overall larger than the observation on a large scale. On a small scale, there is a shape difference as no beam function (smoothing) is applied to the simulation. The ratio between two power spectra is similar to the ratio between two cosmologies at $\ell<100$. To correct for the difference from a number of sources, we adopt a simple empirical model with transfer function and white noise terms as follows:
\begin{equation}
    \label{eq:Eq3}
    C_{\ell}^{obs} = C_{\ell}^{mock} = b_{\ell}^2 (C_{\ell}^{sim} + N_{\ell,1}) + N_{\ell,2}
\end{equation}
where $C_{\ell}^{obs}$ is the angular power spectrum of the reconstructed CIB map, $C_{\ell}^{mock}$ is the readjusted angular power spectrum of the mock CIB, and $C_{\ell}^{sim}$ is the raw angular power spectrum from MXXL. There are three fitting terms: $b_{\ell}$ is the transfer function accounting primarily for the beam smearing on small scales plus minor cosmology rescaling on large scales, and $(N_{\ell,1},N_{\ell,2})$ are the white noise terms introduced before and after the transfer function respectively. $N_{\ell,1}$ can be thought of as a shot noise difference, and the $N_{\ell,2}$ term corresponds to a pure white noise. The underlying assumption of this approach is that the $C_\ell$ match on large scales by construction because $b\,dI/dz$ is scaled to match C23. There should be only a slight difference in amplitude arising from the difference in cosmology. We iteratively find a solution of $(b_\ell, \sigma_{N_1}, \sigma_{N_2})$ to match one-point and two-point statistics simultaneously, where $\sigma_{N_1}$ and $\sigma_{N_2}$ denote the per-pixel standard deviation of the Gaussian white noise terms $N_{\ell,1}$ and $N_{\ell,2}$ respectively. The iteration process is as follows:
\begin{enumerate}
    \renewcommand{\theenumi}{\roman{enumi}}   
    \renewcommand{\labelenumi}{(\theenumi)}   
    \item Initial solution of $(\sigma_{N_1}, \sigma_{N_2})=(0,0)$ is assigned.
    \item The transfer function $b_\ell$ is computed via Equation \ref{eq:Eq3} (i.e., $b_\ell=\sqrt{\frac{C_{\ell}^{obs}-N_{\ell,2}}{C_{\ell}^{sim}+N_{\ell,1}}}$).
    \item Scale the transfer function to be $b_{\ell}\approx\frac{\mathrm{Planck}}{\mathrm{MXXL}}$ at specific $\ell=20$ to match the power spectrum on a large scale. We fit the $b_\ell$ to an exponential functional form of $A\exp(-a\cdot\ell^b)$ to suppress noisy features in the transfer function. In the case of a Gaussian beam, $b\approx2$ and $\sigma\approx\sqrt{2a}$.
    \item Find the best $(\sigma_{N_1}, \sigma_{N_2})$ that matches the probability distribution function (PDF) of pixel intensity values.
\end{enumerate}
These steps are repeated until $(\sigma_{N_1}, \sigma_{N_2})$ converges so that both $C_{\ell}$ and the PDF of C23 align well with those of mock data. To add white noise at the map level, we assign a Gaussian random variable ${\sim} N(0, \sigma^2)$ to each pixel. We find that $(\sigma_{N_1}, \sigma_{N_2})=(0, 0.02)\,\mathrm{MJy}~\mathrm{sr}^{-1}$ gives the best fit to the observed $C_\ell$ and PDF.

In Figure \ref{fig:Fig4}, we show the angular power spectrum and PDF of our readjusted mock data. The mock data (blue dashed lines) align with the observational data (black lines) within the uncertainties (gray shading, estimated from 200 realizations).
The red dotted line shows the case where the simulated map is smoothed with the transfer function defined by $b_\ell=\sqrt{C_{\ell}^{obs}/C_{\ell}^{sim}}$. While $C_\ell$ is tuned to fit the observation, noise terms are essential to match the shape of the PDF. 

\begin{figure*}[t]
\centering
\includegraphics[width=0.99\textwidth]{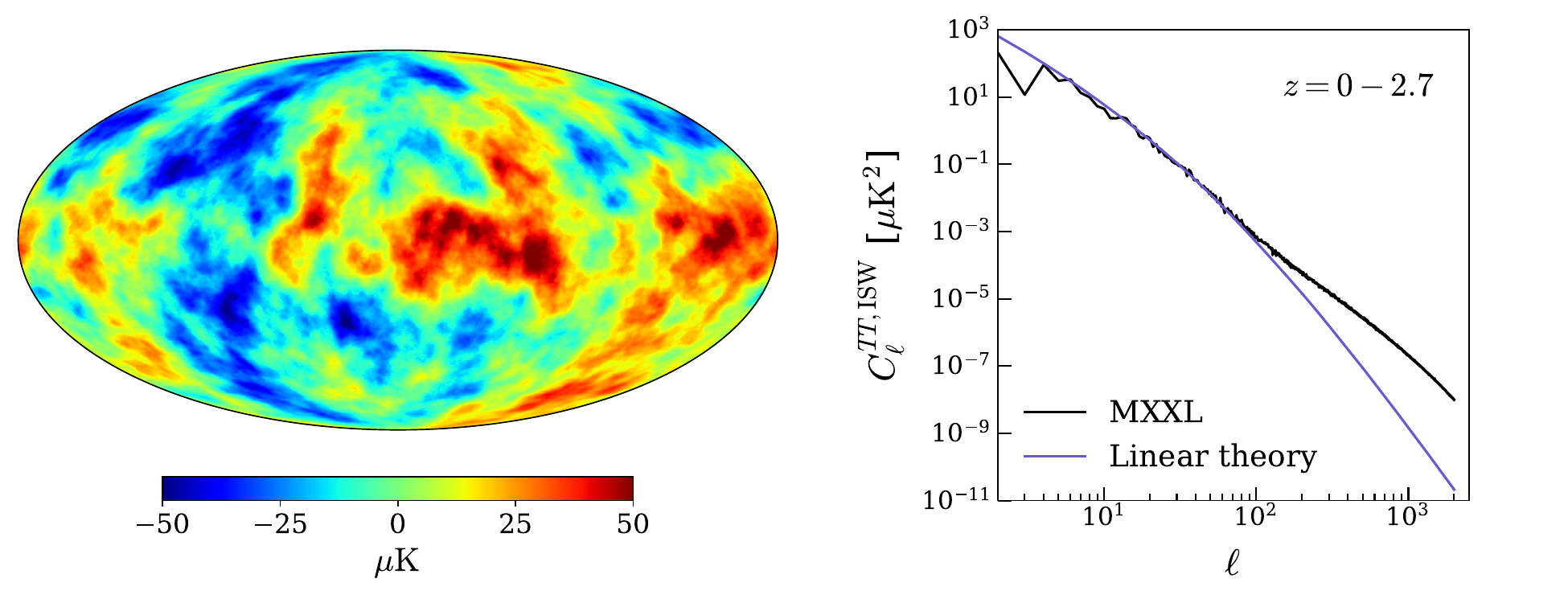}
\caption{Left: mock ISWRS map ($z=0\text{--}2.7$) constructed from the density fields of MXXL simulation. Right: angular power spectrum of the mock ISWRS map (black) compared with the power spectrum in the linear theory (blue). The mock is consistent with linear theory on a large scale ($\ell\lesssim100$).
\label{fig:Fig6}}
\end{figure*}

To validate the mock data, we compute the PDFs of top-hat smoothed fields, which are constructed from information in nearby pixels (the $k$ nearest neighbors for discrete data). The top-hat smoothed fields in turn contain the information of higher-order statistics, which is the key ingredient of $k$NN-CDF cross correlation. In Figure \ref{fig:Fig5}, we present the PDFs of the CIB maps smoothed with a top-hat filter at radii $\theta=0\fdg1,0\fdg5,$ and $0\fdg9$. The measurements from the mock data (blue dashed lines) agree with the C23 map (black solid lines) within $2\sigma$ calculated with 200 realizations of the CIB. Although the mock data are tuned to match only one-point (PDF) and two-point statistics ($C_{\ell}^{\mathrm{CIB-CIB}}$), the PDFs of smoothed fields overall match with the observation, indicating that the integral of higher-order statistics also matches.\\

\subsubsection{Mock ISWRS map \label{sec:Sec2.2.2}}
To simulate temperature fluctuations due to the ISWRS effect, we calculate the evolution of the gravitational potential in the MXXL simulation, as shown in the left panel of Figure \ref{fig:Fig6}.
For a line of sight $\hat{n}$, the CMB temperature variation induced by the ISWRS effect is calculated as follows:
\begin{equation}
\label{eq:Eq4}
\begin{aligned}
    \Delta T_{\mathrm{ISWRS}}(\hat{n}) &= \frac{2}{c^3} \overline{T}_{\mathrm{CMB}} \int \dot{\Phi}(r,\hat{n})~a~dr\\
    &= \frac{2}{c^2} \overline{T}_{\mathrm{CMB}} \int \frac{d\Phi}{da}(z,\hat{n})~a^2~dz 
\end{aligned}
\end{equation}
where $\overline{T}_{\mathrm{CMB}}=2.7260\pm0.0013\,\mathrm{K}$ \citep{Fixsen2009} is the mean CMB temperature, $\dot{\Phi}$ is the time derivative of the gravitational potential, and $a$ is the scale factor. MXXL simulation provides particle counts in a $1024^3$ regular grid at 51 snapshots ranging from $z\approx9.3$ to $z=0$, irregularly spaced with $\Delta a \approx 0.02$. The time derivative of the gravitational potential can be calculated using density and momentum fields in Fourier space. However, as momentum fields are not available in MXXL simulation, we use cubic spline interpolation in scale factor $a$ space to calculate $d\Phi/da$ at a target scale factor $a_{\mathrm{target}}$. 
We calculate the potential field at each time step using the Poisson equation in Fourier space, $\Phi(k,t) = \frac{3}{2} (\frac{H_0}{k})^2 \Omega_{m,0} \frac{\delta(k,t)}{a}$. The corresponding potential field in real space can be obtained by inverse Fourier transformation. 
To align the density field with the lightcone halo catalog, we place the observer at the same position as in the previous section: $(x,y,z)=(17,163,2771)\,h^{-1}\,\mathrm{Mpc}$. We then calculate the distance from the observer to every grid point in a $4096^3$ mesh obtained by tiling $4^3$ simulation boxes, and interpolate the lightcone $d\Phi/da$.\footnote{While only $3^3$ boxes are required to cover our redshift range (i.e., $z=0-2.7$), $4^3$ boxes are needed as the observer is not located at the center of the periodic box.} 

We set the line-of-sight direction ($\hat{n}$) using $N_{\mathrm{side}}=2048$ HEALPix and integrate $d\Phi/da$ at $(z,\hat{n})$ up to $z=2.7$ with radial spacing of $\Delta z=0.001$. We use cloud-In-cell (CIC) interpolation\footnote{Using nearest grid point (NGP) or triangular shaped cloud (TSC) interpolation alters the final result by less than 1\%.} \citep{Hockney1988}, or trilinear interpolation using the eight adjacent grid points, to calculate $\frac{d\Phi}{da}$ at a random position in the lightcone space.
The right panel of Figure \ref{fig:Fig6} shows the angular power spectrum of the resulting ISWRS map together with the linear theory prediction in the same redshift range. We calculate the power spectrum in the linear theory as follows:
\begin{multline}
\label{eq:Eq5}
C_{\ell}^{\mathrm{ISW-ISW}}
= \frac{9H_{0}^{4}\Omega_{m,0}^2\overline{T}_{\mathrm{CMB}}^2}{(\ell+1/2)^{4}c^{6}}
\int dr\, r^2 \\
\qquad \times H^{2}(1-\beta)^2 \, P\left( k=\frac{\ell+1/2}{r},z \right)
\end{multline}
where $\beta\equiv\frac{d \ln D}{d \ln a}$ is the linear growth rate ($D$: linear growth factor) and $P(k,z)$ is the matter power spectrum at redshift $z$. The resulting ISWRS anisotropy is dominated by large angular scales (low multipoles),
while small-scale nonlinear contributions are subdominant \citep[see also][]{Cai2009}.\\

\section{Methods \label{sec:Sec3}}
\subsection{Cross Correlations with $k$NN-CDF \label{sec:Sec3.1}}
The $k$NN-CDF quantifies cross correlation as the joint probability of finding nearby neighbors within a given volume in two discrete point sets such as galaxies and dark matter halos \citep{Banerjee2021B}. \citet{Banerjee2023} have extended this framework to cross correlation between a discrete field and a continuous field: the continuous field is smoothed with a top-hat filter of radius $\theta$, and pixels exceeding a threshold are counted, analogous to finding more than $k$ neighbors in discrete data. We apply this idea to cross-correlate two continuous fields by defining
\begin{equation}
    \label{eq:Eq6}
    \psi_{T^*, I^*}(\theta) \equiv
    \dfrac{P_{> T^* , > I^*}}{P_{> T^*} \times P_{> I^*}} - 1\,,
\end{equation}
where $\theta$ is the angular smoothing scale, $T^*$ and $I^*$ are the thresholds for the CMB temperature and $100\,\micron$ intensity maps, and $P$ denotes the fraction of pixels satisfying the subscript condition. This statistic vanishes for uncorrelated fields and is positive (negative) when the two fields are positively (negatively) correlated.

We compute $\psi$ with our datasets following the framework presented by \citet{Gupta2024}. Setting $N_{\mathrm{side}}=2048$, we subtract the monopole from the CMB temperature anisotropy map. For the $100\,\micron$ intensity map, which is nominally monopole-free, we remove any small residual for consistency. We then smooth them in harmonic space

\begin{equation}
\label{eq:Eq7}
a_{\ell m}^{\mathrm{top-hat}} = b_{\ell}\, a_{\ell m}
\end{equation}
by applying the filter
\begin{equation}
b_{\ell} = \frac{1}{2\ell+1}\, \frac{P_{\ell-1}(\cos\theta)-P_{\ell+1}(\cos\theta)}{1-\cos\theta}
\end{equation}
where $a_{\ell m}$ are spherical harmonic coefficients and  $P_{\ell}(\cos\theta)$ is the Legendre polynomial of degree $\ell$. 
The pixels behind the generic mask are zero-filled, and those near the mask are additionally masked after smoothing, as described below. 
Finally, after the smoothing and masking, we count the number of pixels exceeding the given threshold values $T^*$ and $I^*$, respectively, for the CMB and $100\,\micron$ CIB map.

With the smoothed maps, we adopt fixed percentile thresholds of $(T^*,I^*)\in \{ 50\%, 70\%, 90\% \}$ and angular scales of $\theta\in\{0.1\arcdeg, 0.3\arcdeg, 0.5\arcdeg, 0.7\arcdeg, 0.9\arcdeg\}$. The angular grid is bounded above by the sky coverage and below by the angular resolutions of the CMB and CIB maps. As mentioned in Section \ref{sec:Sec2.1.3}, we further mask all pixels within $\theta$ of the existing mask boundary to prevent bias from the map-smoothing step; consequently, the sky fraction $f_{\textrm{sky}}$ decreases rapidly with $\theta$ (e.g., for the generic mask with $f_{\textrm{sky}}\approx0.56$, $f_{\textrm{sky}}$ drops to $\approx0.15$ for a $1\arcdeg$ top-hat filter and to $<0.01$ for a $2\arcdeg$ filter). We therefore focus on subdegree scales. On the smallest scales, the SFD map has an effective beam size of $\mathrm{FWHM}=6\arcmin$, and the Planck temperature map has an approximate resolution of $\mathrm{FWHM}=5\arcmin$, which motivates our choice of $\theta_{\min}=0.1\arcdeg$. Accordingly, we use multipoles up to $\ell_{\max}=2000$ when smoothing the maps using Equation \ref{eq:Eq7}.

The preprocessing (smoothing, masking) and pixel counting ($>T^*,I^*$) steps described above are applied to three different datasets. The first is the cross correlation between the C23 CIB map and Planck CMB temperature map, which comprises our data vector. The second consists of the random vectors used for covariance estimation, composed of 200 measurements from 200 realizations of CIB maps and random CMB maps. The random CMB maps are generated using the Planck best-fit power spectrum to $\Lambda$CDM. We apply $\mathrm{FWHM}=5\arcmin$ Gaussian smoothing to mimic the Planck temperature map. We also adopt the noise maps from full-focal-plane simulations \citep[FFP10;][]{Planck2016FFP}\footnote{\url{https://pla.esac.esa.int}} and add to the random CMB. The last is the mock vector, where we calculate $\psi$ with mock CIB and ISWRS maps, but with 1000 random CMB maps added to the ISWRS map. In summary, each $\psi$ has $3\times3=9$ combinations of thresholds and five angular scales, resulting in a $9\times5=45$ element vector. The random and mock datasets have 200 and 1000 realizations of $\psi$ respectively.\\

\subsection{Covariance Matrix and Signal-to-noise Ratio \label{sec:Sec3.2}}

We estimate the covariance matrix using the random vectors
\begin{equation}
    \label{eq:Eq9}
    C_{ij} = \frac{1}{N-1} \sum (\psi_i - \overline{\psi}_i)(\psi_j - \overline{\psi}_j)
\end{equation}
where $\psi$ are the random vectors, $(i,j)=1,\dots,45$ is the index within a random vector, and $N=200$ is the number of random vectors. The inverse of the covariance matrix is corrected using the Hartlap factor described by \citet{Hartlap2007}. The Hartlap factor for our case is $\frac{200-45-2}{200-1}\approx0.77$.

The detection significance is estimated with two quantities. The first is the deviation from the null hypothesis to assess the existence of the correlation signal. $\chi_{\textrm{null}}^2$ is defined as follows:
\begin{equation}
    \label{eq:Eq10}
    \chi_{\textrm{null}}^2 = (\psi_{obs} - \overline{\psi}_{random})^{\mathsf T}\,C^{-1}\,(\psi_{obs} - \overline{\psi}_{random})
\end{equation}
We subtract the mean of random vectors to take into account any systematic offset or accidental correlation between CIB and primordial CMB, which is expected to vanish in the absence of systematics.
The second is defined using the best-fit amplitude $A_{k\mathrm{NN}}$ compared to the mock data. An amplitude of $A_{k\mathrm{NN}}=1$ indicates that the ISWRS signal in the data matches the $\Lambda$CDM prediction, whereas $A_{k\mathrm{NN}}=0$ implies no detectable ISWRS signal relative to that prediction. Thus, we estimate the S/N by $A_{k\mathrm{NN}}/\sigma_{A_{k\mathrm{NN}}}$, where $\sigma_{A_{k\mathrm{NN}}}$ is the uncertainty assigned for the amplitude estimation.
The best-fit amplitude and the corresponding uncertainty are estimated by minimizing
\begin{equation}
    \label{eq:Eq11}
    \chi^2 = (\psi_{obs} - A_{k\mathrm{NN}} \cdot \overline{\psi}_{model})^{\mathsf T}\,C^{-1}\,(\psi_{obs} - A_{k\mathrm{NN}} \cdot \overline{\psi}_{model})    
\end{equation}
where the model is any assumed cosmological model---in our case $\Lambda \mathrm{CDM}$---and $\overline{\psi}_{model}$ is the mean of the $\psi$ measured with the mock data. $A_{k\mathrm{NN}}$ and the variance $\sigma_{A_{k\mathrm{NN}}}^2$ can be determined by differentiating $\chi^2$ with respect to $A_{k\mathrm{NN}}$
\begin{equation}
    \label{eq:Eq12}
    A_{k\mathrm{NN}} = \frac{\psi_{model}^{\mathsf T}\,C^{-1}\,\psi_{obs}}{\psi_{model}^{\mathsf T}\, C^{-1}\,\psi_{model}}
\end{equation}

\begin{equation}
    \label{eq:Eq13}
     \sigma_{A_{k\mathrm{NN}}}^2 = \left[ \psi_{model}^{\mathsf T}\,C^{-1}\,\psi_{model} \right]^{-1}
\end{equation} 

\begin{equation}
    \label{eq:Eq14}
     \mathrm{S/N} = \frac{A_{k\mathrm{NN}}}{\sigma_{A_{k\mathrm{NN}}}}= \sqrt{\chi_{\textrm{null}}^2-\chi_{\textrm{best--fit}}^2}
\end{equation}
\\

\subsection{Mask Selection and Outlier Treatment \label{sec:Sec3.3}}

\subsubsection{Sky Area for Convergence Tests \label{sec:Sec3.3.1}}

\begin{figure*}[t]
\centering
\includegraphics[width=0.99\textwidth]{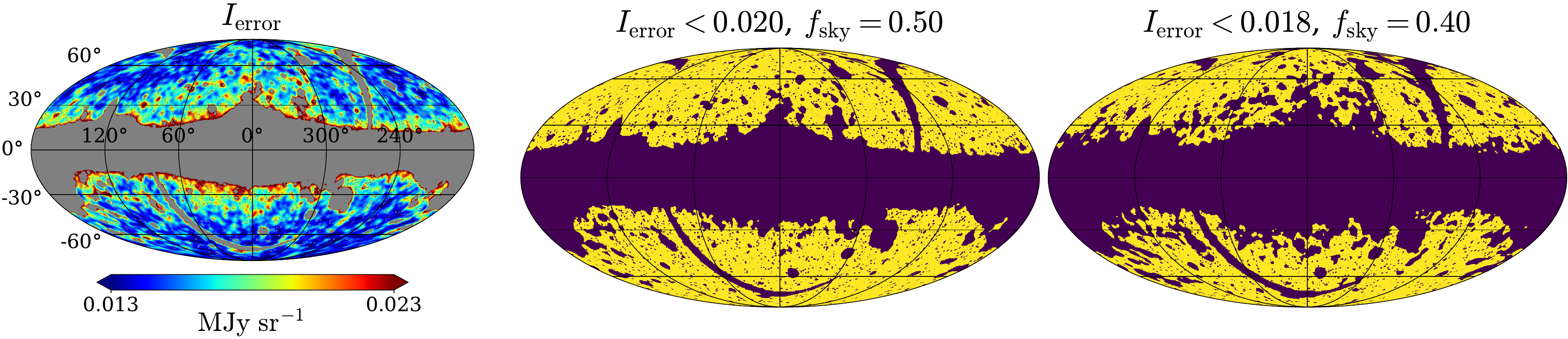}
\caption{Left: CIB reconstruction error $I_{\mathrm{error}}\,(\mathrm{MJy}~\mathrm{sr}^{-1})$ in Galactic coordinates. The original error map from C23 is smoothed with an $\mathrm{FWHM}=1\arcdeg$ Gaussian to trace overall uncertainty structure. Middle, right: two examples of masks constructed with the smoothed error map (left panel). In addition to the generic mask, regions with $I_{\mathrm{error}}$ larger than $(0.02, 0.018)\mathrm{MJy}~\mathrm{sr}^{-1}$ are masked.
\label{fig:Fig7}}
\end{figure*}

In Section \ref{sec:Sec2.1.3}, we constructed the generic mask by combining the common mask of the Planck CMB temperature map and the CIB mask that is based on the reconstruction process. However, the reliability of the CIB map still depends on the Galactic latitude, or effectively the reconstruction error (see the left panel of Figure \ref{fig:Fig7}).
The error arises from multiple effects in the template galaxy density fields, including dust extinction and its correction, source detection in crowded fields, and star--galaxy separation, all of which trace Galactic structure.
While a straightforward approach is to apply a Galactic latitude or $E(B-V)$ cut, we instead construct a new set of masks based on the CIB reconstruction error map to capture these contributions to the uncertainty. We first smooth the error map with a Gaussian filter to suppress small-scale noisy patterns. We test smoothing scales of $\mathrm{FWHM}= 0\fdg5\text{--}5\arcdeg$ and empirically find that a smoothing scale of $1^{\circ}$ best captures the uncertainty structure. Smaller scales (e.g., $0\fdg5$) retain noisy features, whereas larger scales (e.g., $5\arcdeg$) wash out small-scale detail. Starting from the baseline mask ($f_{\textrm{sky}}\approx0.56$), we progressively enlarge the masked area by excluding regions with large values in the smoothed reconstruction error map; varying the threshold yields masks spanning $f_{\textrm{sky}}\approx0.40-0.55$. The middle and right panels of Figure \ref{fig:Fig7} show two examples of such masks when retaining regions with $I_{\mathrm{error}}<0.02$ and $0.018~\mathrm{MJy}~\mathrm{sr}^{-1}$.

We measure the $k$NN statistics and corresponding covariance matrices across this suite of masks and examine the resulting trends. As we remove more area near the Galactic plane, the purity of the CIB map (and partly the CMB map) is expected to improve, but the statistical power decreases with decreasing $f_{\textrm{sky}}$. Accordingly, we expect $\sigma_{A_{k\mathrm{NN}}}$ to decrease and reach a minimum at an intermediate $f_{\textrm{sky}}$, and then increase if substantial contamination remains in the CIB map. The measured-to-theoretical ratio $A_{k\mathrm{NN}}$ may also vary as noise and residual contamination suppress the true signal. Ultimately, our cosmological interpretation should be robust under our adopted masking choice. In Section \ref{sec:Sec4}, we present the results as a function of $f_{\textrm{sky}}$ and define our fiducial mask used for the final measurements. \\

\subsubsection{Outlier Removal: Blind Test with the Mock Data \label{sec:Sec3.3.2}}
Previous $k$NN-CDF analyses typically exclude the PDF tails. While the tails encode information about rare extrema, they are also more susceptible to statistical noise \citep[e.g.,][]{Banerjee2021B}. This choice is particularly important in our case because the signal is embedded in noise that is much larger than the signal. A common convention is to remove the top 5\% of pixels, which roughly corresponds to a one-sided $2\sigma$ cut; the robustness of this choice has been explored in earlier work \citep[e.g.,][]{Anbajagane2023}.

\begin{figure}[t]
\centering
\includegraphics[width=0.85\linewidth]{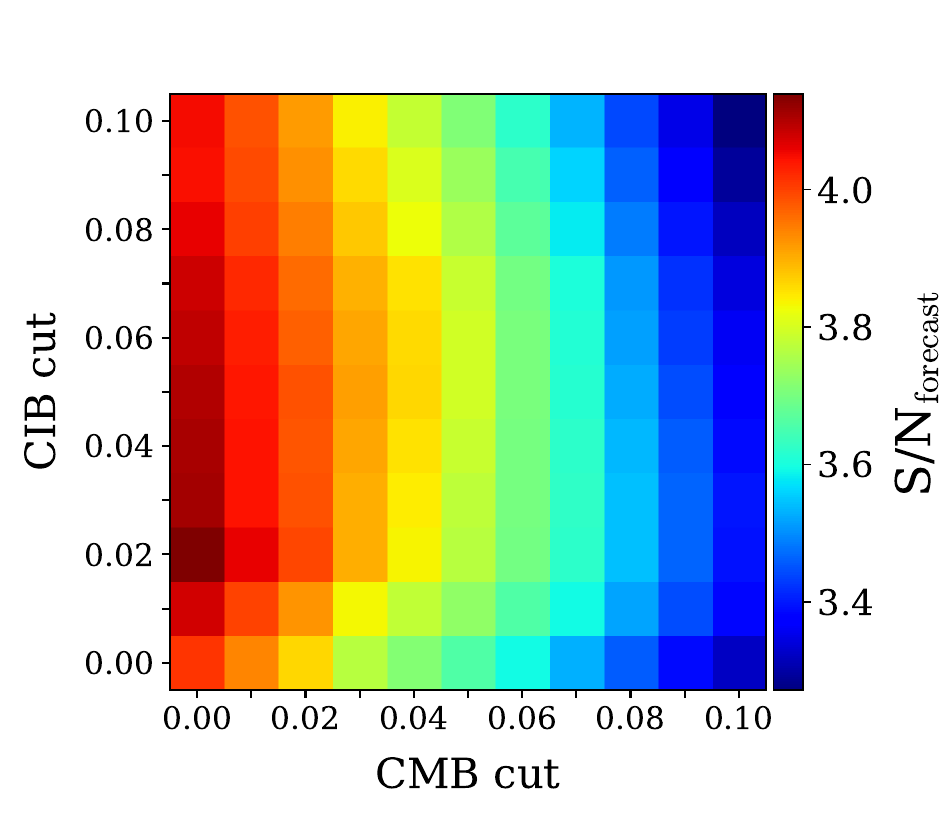}
\caption{Expected S/N calculated with the mock data. Each pixel denotes a different combination of outlier cuts (CMB cut, CIB cut). The S/N peaks at $(0,0.02)$ and decreases when any CMB outlier cuts or larger CIB outlier cuts are applied. The variation is dominated by the direction of the CMB cut.
\label{fig:Fig8}}
\end{figure}

Using the mock data, we assess the sensitivity to outlier cuts around the conventional 5\% threshold and adopt a fiducial cut for subsequent analyses. Specifically, we shift the percentile thresholds (50\%, 70\%, 90\%, and 100\%) by 0–10\%. Figure \ref{fig:Fig8} shows the forecasted S/N for these choices: within the tested range, it is highest for $(0, 0.02)$ and is ${\sim}26\%$ larger than in the lowest-S/N case, $(0.1, 0.1)$. Because the outlier cut can influence both the interpretation and the robustness of the measurement, we emphasize two limitations. First, outlier cuts can alter the effective redshift sensitivity of the $k$NN-CDF. For example, the brightest peaks in our CIB map are predominantly associated with nearby clusters, so trimming the high-end tail partly shifts the redshift range emphasized by the statistic. The fiducial choice adopted here is therefore not universal and should be reassessed for other fields. Second, our mock data do not include the reconstruction errors present in the observational data. Although we mask the lowest-latitude regions with the largest errors using the masks constructed in Section \ref{sec:Sec3.3.1}, spurious noise-induced peaks remain. We therefore use $(0, 0.02)$ as our fiducial choice, while also examining the full range of cuts in $[0, 0.05]$ to quantify their impact on the final results.\\

\section{Results \label{sec:Sec4}}
In this section, we present our $A_{k\mathrm{NN}}=0.95\pm0.20\,(4.8\sigma)$ detection of the ISWRS effect using a mask with $f_{\textrm{sky}}\approx0.49$. We first present the observed cross-correlation signal (Figure \ref{fig:Fig9}) and its deviation from the null hypothesis. We then compare the measured $k$NN signal with the $\Lambda$CDM mock data (Figure \ref{fig:Fig13}) to determine the best-fit amplitude $A_{k\mathrm{NN}}$.

\subsection{Measured Cross-correlation Signal \label{sec:Sec4.1}}

\begin{figure*}[t]
\centering
\includegraphics[width=0.99\textwidth]{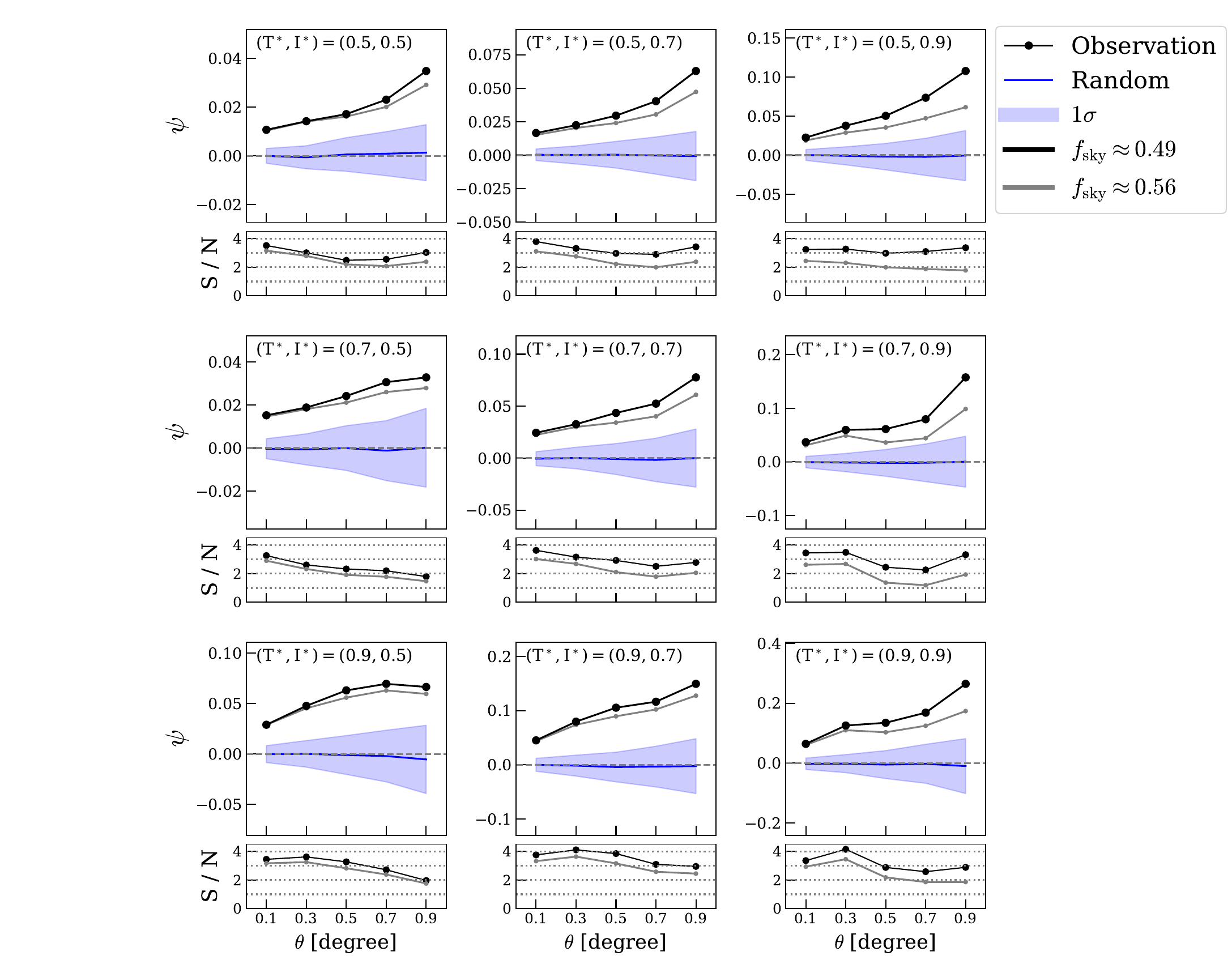}
\caption{Top: $k$NN cross-correlation statistic $\psi$ as a function of angular scale $\theta$. Each panel shows a different combination of thresholds (text on the upper left). Measurements from the observational data (black lines) are shown together with the distribution of random vectors (blue lines). The blue shading denotes the standard deviation of the random vectors. Bottom: deviation from the null hypothesis for a single bin. Even with a single bin, S/N reaches up to $4.1\sigma$. While the random vector (blue) in this figure is computed using the $f_{\textrm{sky}}\approx0.49$ mask, the calculation with the $f_{\textrm{sky}}\approx0.56$ mask (gray) is shown together for comparison.
\label{fig:Fig9}}
\end{figure*}

\begin{figure}[t]
\centering
\includegraphics[width=0.99\linewidth]{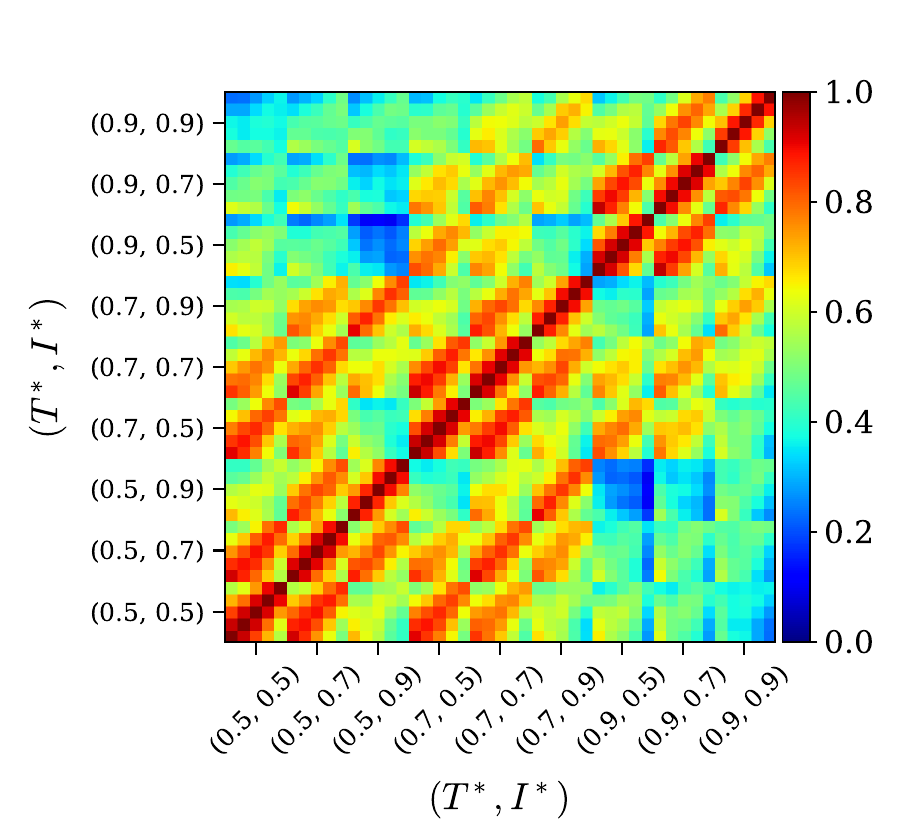}
\caption{Correlation matrix of $k$NN summary statistic $\psi$ defined as $C_{ij} / \sqrt{C_{ii}C_{jj}}$ (see Equation \ref{eq:Eq9}). Each $5\times5$ square shows the correlation between two pairs of $(T^*,I^*)$, and the five bins within that square show the correlation between angular scales. As $\psi$ is defined in real space, angular bins are highly correlated up to $0.93$. Different combinations of $(T^*,I^*)$ are also correlated, as a cumulative distribution is used in this method.
\label{fig:Fig10}}
\end{figure}

\begin{figure}[t]
\centering
\vspace{7mm}
\includegraphics[width=0.99\linewidth]{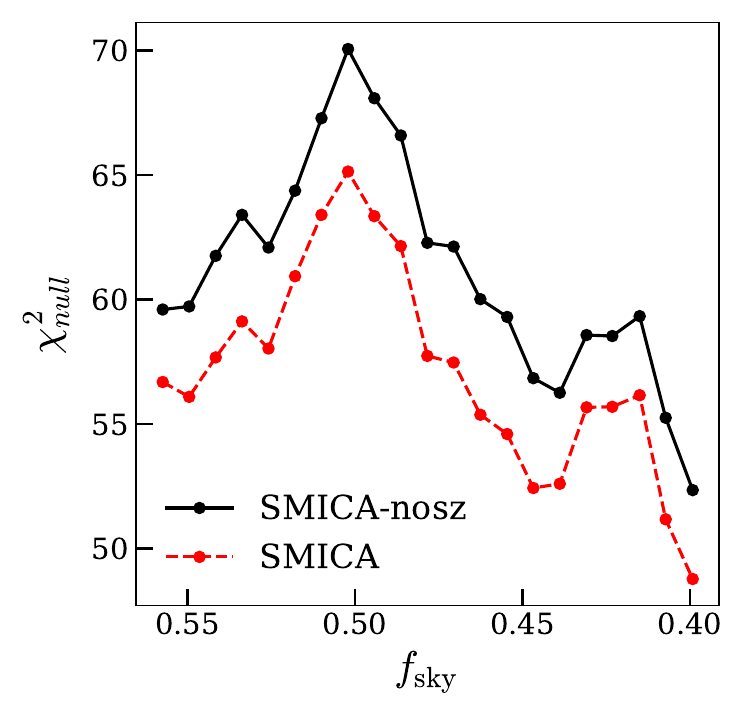}
\caption{$\chi_{\textrm{null}}^2$ as a function of $f_{\textrm{sky}}$. $\chi_{\textrm{null}}^2$ increases up to $\approx70$ when $\approx6\%$ of the original map is additionally masked. The SMICA map without removal of the SZ effect shows a lower signal overall.
\label{fig:Fig11}}
\end{figure}

Figure \ref{fig:Fig9} shows the measured $k$NN cross-correlation statistic $\psi$ as a function of angular scale $\theta$. In this figure, we apply a $2\%$ outlier cut on the CIB according to the analysis in Section \ref{sec:Sec3.3.2}. Each panel shows one combination of thresholds (text on the upper left) used for the measurement. Black data points are the measurements from the observational data (i.e., the Planck temperature map and the $100\,\micron$ CIB map), and blue lines denote the mean of 200 random vectors.
The random vectors are consistent with zero, showing no evidence for significant systematics or accidental cross correlation between the CIB and primary CMB. The shading around the mean of random vectors shows $1\sigma$ (the standard deviation) of the random vector distribution. The corresponding significance for individual bins is shown in the bottom panels. Individual bins deviate ($1.8\sigma$--$4.1\sigma$) from the null hypothesis, especially reaching higher significance for smaller angular scales where the sky coverage is larger. The data vector and the corresponding S/N (gray) with the $f_{\textrm{sky}}\approx0.56$ mask are shown together for comparison.

As $\psi$ is calculated in real space and is based on a CDF, measurements are highly correlated between different angular scales and thresholds. Figure \ref{fig:Fig10} shows the correlation matrix estimated with the random vectors. Each $5\times5$ shape square represents the correlation between two $(T^*,I^*)$ pairs, and the five components in each square show the correlation between angular bins. The angular bins are highly correlated up to $\approx0.93$. Different combinations of thresholds are also highly correlated, but overall less so than the angular scales.

Figure \ref{fig:Fig11} shows $\chi_{\textrm{null}}^2$ as a function of $f_{\textrm{sky}}$, accounting for the correlation shown in Figure \ref{fig:Fig10}. Deviation from the null hypothesis increases up to $\chi_{\textrm{null}}^2\approx70~(2.6\sigma)$ at $f_{\textrm{sky}}\approx0.5$ and decreases as a more aggressive mask is applied. In this figure, we also present the result when the original SMICA map (no removal of the SZ effect) is used. The corresponding cross-correlation signal is always lower than the SMICA ``SZ-free'' map, indicating that signatures of spectral distortion by hot gas affect the CMB temperature--density cross correlation.\\

\subsection{Amplitude and Significance Relative to the $\Lambda$CDM Mock Data \label{sec:Sec4.2}}

\begin{figure*}[p]
\centering
\includegraphics[width=\textwidth,height=0.46\textheight,keepaspectratio]{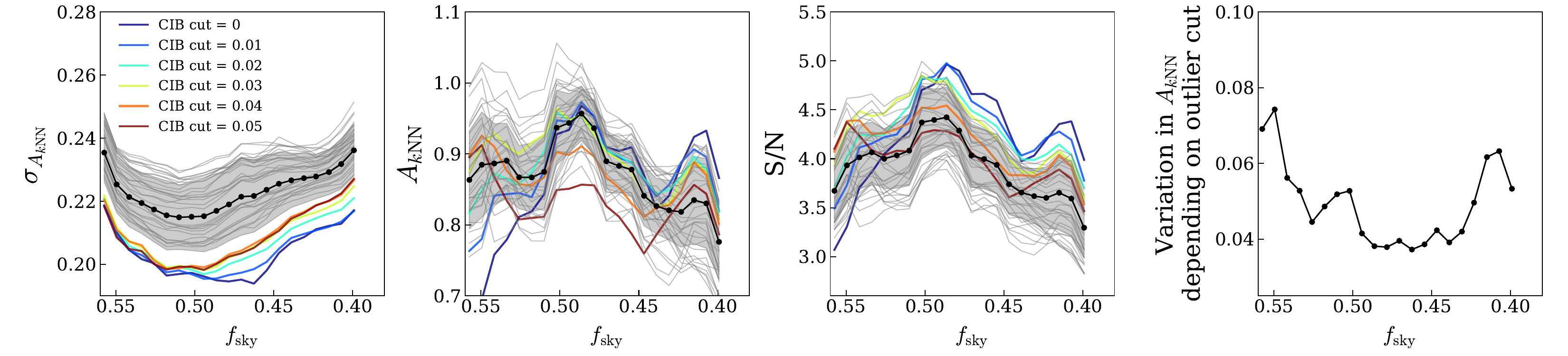}
\caption{Comparison with the $\Lambda$CDM mock data as a function of $f_{\textrm{sky}}$. Different masks constructed in Section \ref{sec:Sec3.3.1} are used to find a robust condition for the detection. The first three panels from the left show $\sigma_{A_{k\mathrm{NN}}}$, $A_{k\mathrm{NN}}$, and S/N. The distribution of the measured quantities for outlier cuts in $[0,0.05]$ is shown with thin gray lines (individual case), black lines (median), and gray shading (standard deviation). Colored lines (from blue to red) denote the case when only the CIB outlier cut is applied. The error reaches a minimum around $f_{\textrm{sky}}\approx0.5$. $A_{k\mathrm{NN}}$ fluctuates substantially when the original mask is used (leftmost data point), such that a $1\%$ modification of the CIB outlier cut results in a $10\%$ variation. The variation as a function of $f_{\textrm{sky}}$ is shown in the rightmost panel. The amplitudes become stable at $f_{\textrm{sky}}\lesssim0.5$. We adopt $f_{\textrm{sky}}\approx0.49$.
\label{fig:Fig12}}
\end{figure*}

\begin{figure*}[p]
\centering
\includegraphics[width=0.99\textwidth]{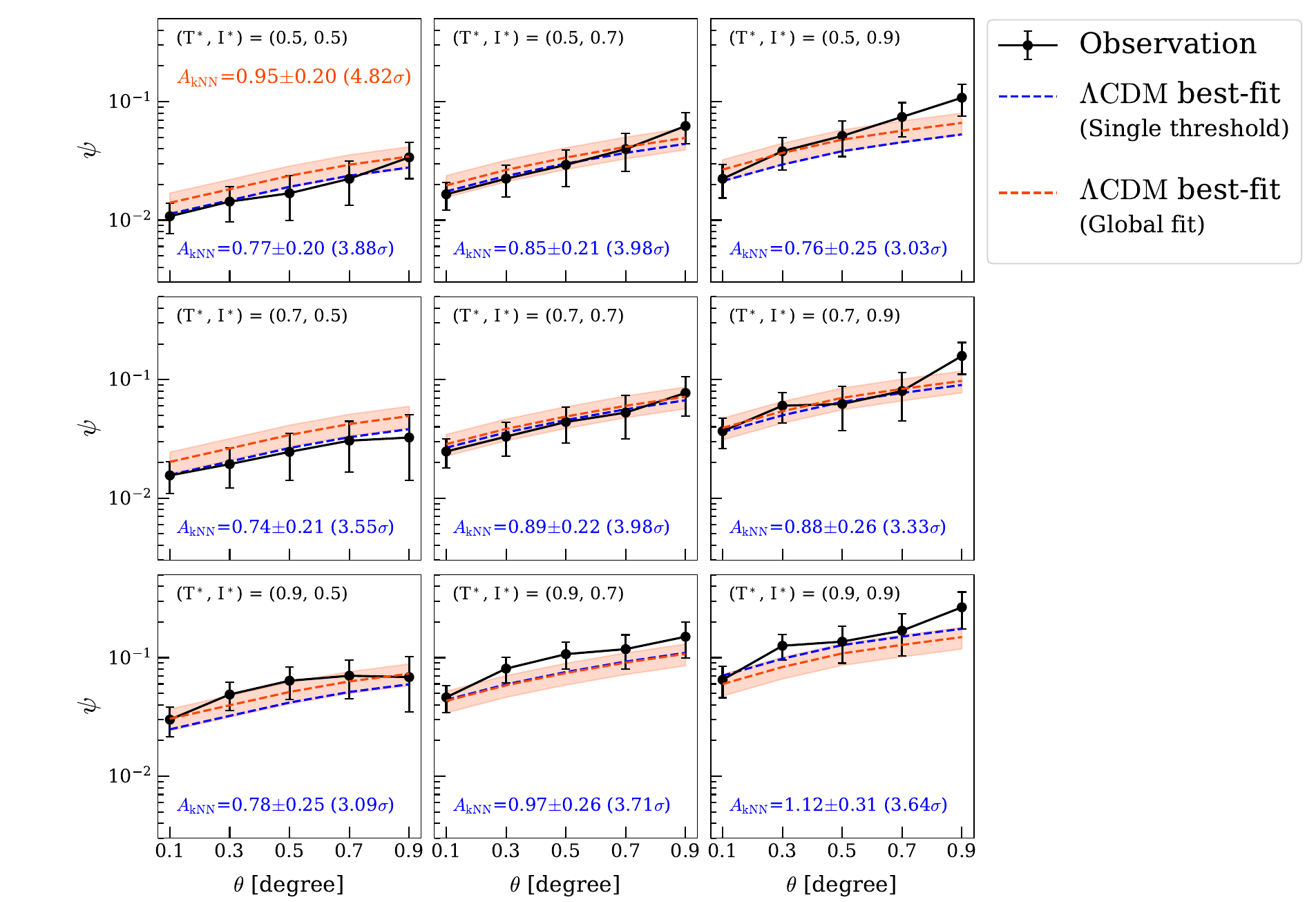}
\caption{$k$NN summary statistic $\psi$ for observation (black markers) and best fit to the $\Lambda$CDM mock data (dashed lines). The blue lines and text are individual estimates for one combination of thresholds $(T^*, I^*)$. The orange lines show the global best fit and the shading shows the $1\sigma$ range of the best-fit amplitude. Each combination of thresholds yields $A_{k\mathrm{NN}}=0.74\text{--}1.12~(3.0\sigma\text{--}4.0\sigma)$. The global fit using the full covariance matrix is $A_{k\mathrm{NN}}=0.95\pm0.20$ ($4.8\sigma$).
\label{fig:Fig13}}
\end{figure*}

To determine the mask to be used, we estimate the best-fit amplitudes using the masks constructed in Section \ref{sec:Sec3.3.1}. The left three panels of Figure \ref{fig:Fig12} show $A_{k\mathrm{NN}}$, $\sigma_{A_{k\mathrm{NN}}}$, and S/N as a function of $f_{\textrm{sky}}$. In all panels, the thin gray lines in the background show the results when different outlier cuts are applied (see Section \ref{sec:Sec3.3.2}). Cases with only the CIB outlier cut applied are color-coded by the cut value (e.g., dark blue: 0, red: 0.05). The black lines and gray shading show the median and standard deviation of the gray lines, namely the distribution of measured values when various outlier cuts are applied.
As the highly uncertain regions are excluded from the analysis, $\sigma_{A_{k\mathrm{NN}}}$ first decreases, reaches a minimum around $f_{\textrm{sky}}\approx0.5$, and then increases. $A_{k\mathrm{NN}}$ and S/N follow a similar trend, increasing up to a certain $f_{\textrm{sky}}$ and then decreasing.

One noticeable trend is the variation in $A_{k\mathrm{NN}}$ and S/N when different outlier cuts are applied. For $f_{\textrm{sky}}\gtrsim0.5$, they fluctuate substantially depending on the outlier cut. The best-fit $A_{k\mathrm{NN}}$ can vary with the outlier cut because it changes the effective redshift range and removes noisy peaks. However, in our data, a 1\% modification of the CIB outlier cut produces up to a 10\% variation (colored lines) in $A_{k\mathrm{NN}}$: specifically, $A_{k\mathrm{NN}}=0.67,0.76,0.82,0.87,0.90$ for $\mathrm{CIB~cut}=0, 0.01, 0.02, 0.03, 0.04$. Although reconstruction error is not included, the corresponding shift in $A_{k\mathrm{NN}}$ is well below $5\%$ in the mock data. This indicates that the tail of the CIB distribution is very sensitive to reconstruction error. Conversely, the variation decreases significantly when the uncertain regions are removed (e.g., $f_{\textrm{sky}}\approx0.49$).
The rightmost panel of Figure \ref{fig:Fig12} shows this variation in $A_{k\mathrm{NN}}$ (i.e., the width of the gray shading in the second panel) as a function of $f_{\textrm{sky}}$. The dispersion rapidly drops by $40\%~(0.07\rightarrow0.04)$ from $f_{\textrm{sky}}\approx0.56$ to $f_{\textrm{sky}}\approx0.49$. Thus, we choose $f_{\textrm{sky}}\approx0.49$ as our final mask, considering both the variation in $A_{k\mathrm{NN}}$ and $\sigma_{A_{k\mathrm{NN}}}$, which reaches a minimum around $f_{\textrm{sky}}\approx0.5$.

With this mask, we first present the results for each combination of thresholds. The results are shown in Figure \ref{fig:Fig13} as blue text and dashed lines. Each combination exhibits $A_{k\mathrm{NN}}=0.74\text{--}1.12$, generally increasing when higher thresholds are used for the CMB. S/N ranges from $3.0$ to $4.0$, peaking at the $(T^*,I^*)=(0.5,0.7)$ combination. The lowest S/N is at the $(T^*,I^*)=(0.5,0.9)$ combination. Combining all combinations with the full covariance matrix (Figure \ref{fig:Fig10}) yields $A_{k\mathrm{NN}}=0.95\pm0.20~(4.8\sigma)$ with an outlier cut of (CMB cut, CIB cut) $=(0,0.02)$. $\mathrm{S/N}$ varies between 3.9 and 5.0 depending on the outlier cut. Except for the case when large outlier cuts are applied to both the CMB and CIB ($\mathrm{CMB~cut}>0.02$ or $\mathrm{CIB~cut}>0.04$; see also Figure \ref{fig:Fig8}), S/N is higher than $4.5$. The results under various conditions are summarized in Table \ref{tab:tab1}.

We note that the amplitude significance ($4.8\sigma$) and the $\chi_{\textrm{null}}^2\approx66.6$ for 45 degrees of freedom ($2.3\sigma$), both evaluated at our fiducial $f_\mathrm{sky}\approx0.49$, are conceptually distinct measures: the former optimally weights the data vector by the $\Lambda\mathrm{CDM}$ signal template through the inverse covariance, while $\chi_{\textrm{null}}^2$ weights all modes of the data vector equally and is therefore a more conservative measure. Indeed, individual bins deviate from the null hypothesis by up to $4.1\sigma$ (Figure \ref{fig:Fig9}), comparable to the full amplitude significance. The best fit to the mock data gives $\chi_{\textrm{best--fit}}^2=43.3$ for $45$ degrees of freedom, and the improvement over the null hypothesis, $\Delta\chi^2$, quantifies the detection consistently with $A/\sigma_A$. A difference of this magnitude between $A/\sigma_A$ and $\chi_{\textrm{null}}^2$ is typical in our mock realizations, where the offset between two significances is $(2.3\pm0.8)\sigma$, consistent with the observed value. \\

\begin{deluxetable}{lccc}
\tablecaption{$A_{k\mathrm{NN}}$ and S/N at $f_{\mathrm sky}\approx0.49$.\label{tab:tab1}}
\tablecolumns{4}
\tablewidth{0pt}
\tablehead{
\multicolumn{1}{l}{Condition} & \colhead{Outlier Cut} & \colhead{$A_{k\mathrm{NN}}$} & \colhead{S/N}
}
\startdata
Mock                & $(0.00,\,0.02)$    & $0.95\pm0.20$ & $4.82$ \\
Minimum $\sigma_{A_{k\mathrm{NN}}}$ & $(0.00,\,0.00)$    & $0.97\pm0.19$ & $4.97$ \\
Highest S/N         & $(0.00,\,0.01)$ & $0.97\pm0.20$ & $4.98$ \\
No outlier cut      & $(0.00,\,0.00)$       & $0.97\pm0.19$ & $4.97$ \\
\enddata
\tablecomments{``Mock" in the first column denotes the best outlier cut combination determined based on the mock data. $A_{k\mathrm{NN}}$: best-fit amplitude; S/N: detection significance.}
\end{deluxetable} 

\section{Discussion \label{sec:Sec5}}
\subsection{Comparison with Two-point Function Analysis \label{sec:Sec5.1}}
\begin{figure}[t]
\centering
\includegraphics[width=0.99\linewidth]{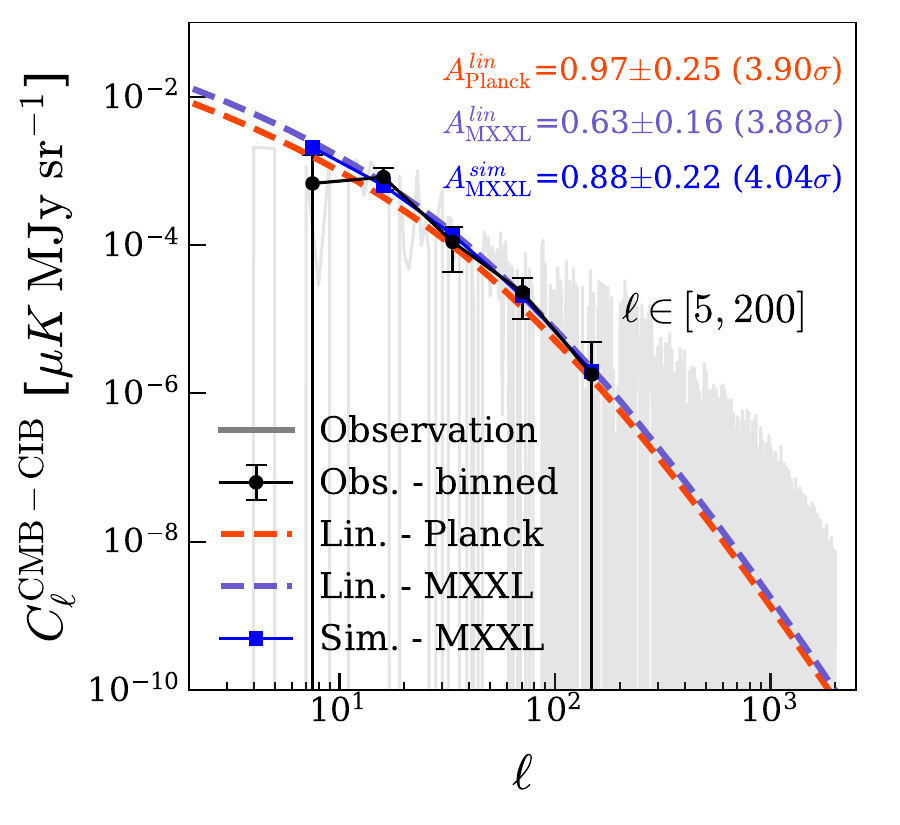}
\caption{Cross angular power spectrum between CMB (ISW) and $100\,\micron$ CIB. The black data points denote the measurements from the observational data. The dashed lines are calculated using linear theory (orange: Planck, light blue: MXXL cosmology assumed). Multipoles in the range $\ell\in[5,200]$ are averaged to five bins equally spaced in log space. The error bars show the diagonal terms of the covariance matrix. For both assumed cosmologies, $\mathrm{S/N}\approx3.9$ with different amplitudes. Using the prediction from MXXL (blue) yields $\mathrm{S/N}\approx4.0$, slightly larger than the linear theory. 
\label{fig:Fig14}}
\end{figure}

We achieved a $4.8\sigma$ detection of the ISWRS effect using the novel $100\,\micron$ CIB map and the $k$NN-CDF statistic. Capturing information from higher-order $N$-point statistics, we expect $k$NN statistics to offer advantages over the classical two-point function analysis. For a quantitative assessment of this point, in this section, we reanalyze the data using the angular power spectrum.

We use the same data and mask ($f_{\textrm{sky}}\approx0.49$) to calculate the cross angular power spectrum $C_{\ell}^{\mathrm{CMB-CIB}}$. The covariance matrix is also estimated using Monte Carlo simulation, namely the cross correlation between 200 realizations of CIB and random CMB. The angular power spectra are calculated with \texttt{anafast} in HEALPix. To account for the mode coupling induced by the partial sky coverage, we deconvolve the mask using the \texttt{NaMaster} \citep{Alonso2019}, binning the power spectrum in $\ell\in[5,200]$ using five bins equally spaced in log space. We exclude the lowest multipoles ($\ell<5$) considering (1) low power of CIB on that scale and (2) large cosmic variance. For a given cosmology and bias-weighted redshift distribution $b\,dI/dz$, the corresponding linear cross power spectrum between ISW and CIB is as follows:
\begin{multline}
\label{eq:Eq15}
C_{\ell}^{\mathrm{ISW-CIB}}
= \frac{3H_{0}^{2}\Omega_{m,0}\overline{T}_{\mathrm{CMB}}}{(\ell+1/2)^{2}c^{4}}
\int dr \left(b \frac{dI}{dz}\right) \\ 
\qquad \times H^{2}(1-\beta)\, P\left( k=\frac{\ell+1/2}{r},z \right)
\end{multline}
We calculate the linear power spectra assuming Planck and MXXL cosmologies. Figure \ref{fig:Fig14} shows the cross angular power spectra calculated with the data and linear theory. The error bars show the diagonal terms of the covariance matrix, which is nearly diagonal due to the choice of binning. The best-fit amplitudes estimated using the full covariance matrix (identical to Equation \ref{eq:Eq12}) are $A_{\mathrm{ISW}}^{\mathrm{Planck}}=0.97\pm0.25$ and $A_{\mathrm{ISW}}^{\mathrm{MXXL}}=0.63\pm0.16$, which both yield $\mathrm{S/N}\approx3.9$. Although the best-fit amplitudes differ by $>1\sigma$ due to the difference in assumed cosmologies (i.e., $\Omega_{m,0}=0.25$ and $\sigma_8=0.9$ for MXXL), both detections are lower than $4\sigma$ significance. If we use the mock data from MXXL for theoretical prediction, the significance slightly increases up to $4.0\sigma$.
The $4.8\sigma$ detection with the $k$NN statistic corresponds to increasing the independent data volume by $(4.8/4.0)^2-1=44\%$ relative to the two-point analysis.
As a sanity check, we repeat the calculation using different covariance estimation methods in Appendix \ref{sec:AppB}. The results are consistent to within ${\sim}7\%$; the significance increases by ${\sim}2\%$ if the CIB reconstruction error is neglected, and the fiducial covariance used in the main text gives the most conservative significance.
Small variations to methodologies such as ($\ell_{\min}, \ell_{\max}$) and the binning scheme slightly change the significance but do not meaningfully increase the S/N.

While achieving $4\sigma$ significance from a single cross correlation is encouraging, our results suggest that the $k$NN statistic is more effective in capturing the $100\,\micron$ CIB--ISW cross-correlation signal than the two-point function. We have checked that this improvement is not simply due to the broader range of scales accessed by the $k$NN statistic. Extending the $C_{\ell}$ analysis to smaller scales ($\ell_{\max}=2000$) does not raise its S/N, consistent with the ISW signal being dominated by large, near-linear scales; conversely, restricting the $k$NN measurement to $5\leq\ell\leq200$ reduces but does not eliminate its sensitivity to the cross correlation. Together, these tests indicate that the $k$NN gain originates from non-Gaussian information rather than from the inclusion of additional Fourier modes. \\

\subsection{Systematics Test}
To investigate potential systematics and foreground contamination, we measure the cross correlation between the CIB and the CMB using a range of CMB maps available in the literature. In addition to the maps from the Planck pipelines (SMICA, NILC, SEVEM, and Commander), we use the local-generalized morphological component analysis (L-GMCA) CMB map \citep{Bobin2016}, derived from Planck 2015 data and nine years of Wilkinson Microwave Anisotropy Probe data. The L-GMCA map is known to be free of any detectable thermal SZ contamination, and can therefore be compared directly with the SMICA-nosz map used in our analysis. Table \ref{tab:tab2} lists the measured amplitudes $A_{k\mathrm{NN}}$ and $A_{\mathrm{ISW}}$ for each pipeline. The L-GMCA map yields a result consistent with our original analysis based on the SMICA-nosz map, whereas the other pipelines exhibit lower amplitudes. As noted in Section \ref{sec:Sec2.1.2}, the SZ effect is known to leave cold-spot residuals in Planck CMB maps. This systematic difference in amplitude can therefore be attributed to these cold spots in the denser regions of the large-scale structure.

\begin{deluxetable}{lccc}
\tablecaption{$A_{k\mathrm{NN}}$ and $A_{\mathrm{ISW}}$ Measured Using the CMB Maps from Different Pipelines. \label{tab:tab2}}
\tablecolumns{5}
\tablewidth{0pt}
\tablehead{
\multicolumn{1}{l}{Pipeline} & \colhead{$A_{k\mathrm{NN}}$} & \colhead{$A_{\mathrm{ISW}}$} & \colhead{Thermal SZ}
}
\startdata
SMICA-nosz\tablenotemark{a}  & $0.95\pm0.20$ & $0.97\pm0.25$ & deprojected   \\
SMICA       & $0.80\pm0.20$ & $0.89\pm0.25$ & \nodata  \\
NILC        & $0.73\pm0.20$ & $0.88\pm0.25$ & \nodata  \\
SEVEM       & $0.62\pm0.20$ & $0.90\pm0.25$ & \nodata  \\
Commander   & $0.63\pm0.20$ & $0.87\pm0.25$ & \nodata  \\
L-GMCA      & $0.94\pm0.20$ & $0.98\pm0.25$ & deprojected   \\
\enddata
\tablecomments{For $A_{\mathrm{ISW}}$, we tabulate the value obtained from linear theory with the Planck cosmology. }
\tablenotetext{a}{Our baseline analysis.}
\end{deluxetable} 

\begin{figure*}[t]
\centering
\includegraphics[width=0.99\linewidth]{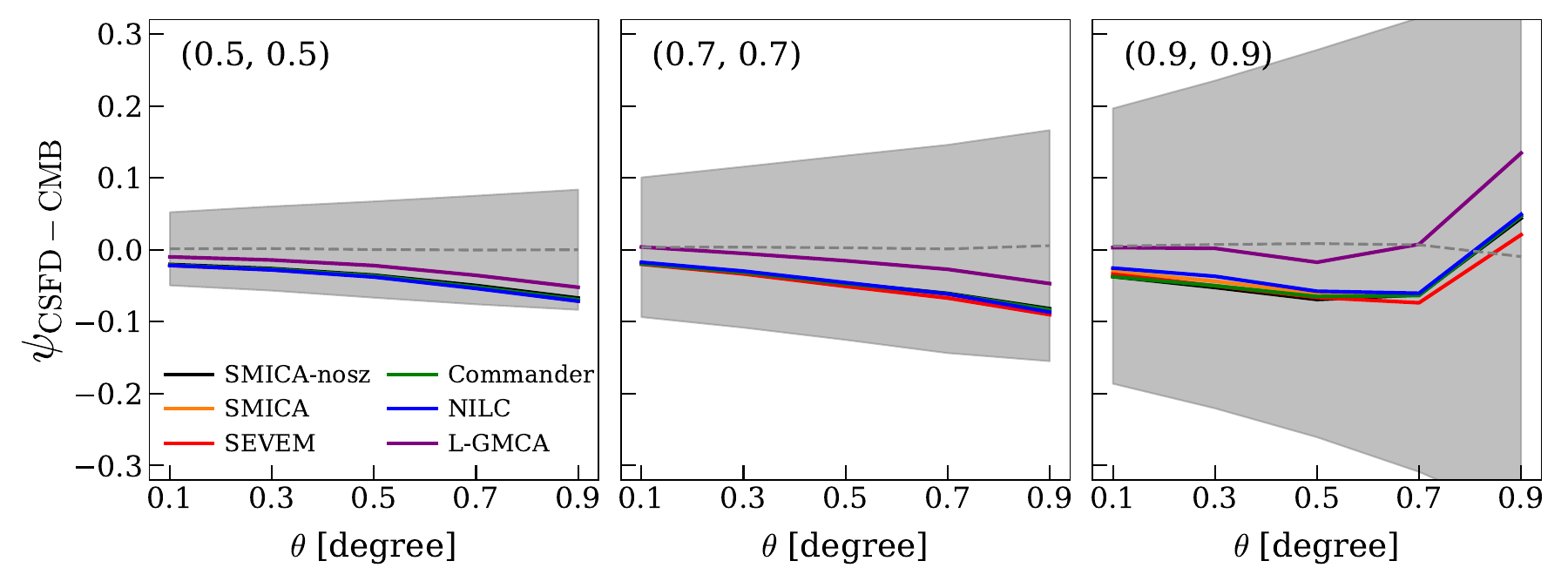}
\caption{$k$NN cross correlation between the CSFD dust map and the CMB maps, $\psi_{\textrm{CSFD-CMB}}$, as a function of angular scale $\theta$. The text on the upper left indicates the threshold combination. The colored lines show the results obtained with the CMB maps from different pipelines. The gray shading shows the standard deviation of the random vectors, estimated from 1000 random CMB maps. Although the measured signal shows a slight negative correlation, it is consistent with the null hypothesis of no correlation.
\label{fig:Fig15}}
\end{figure*}

Besides the SZ residuals discussed above, our results could also be affected by Galactic dust and CIB residuals in the CMB.
To estimate the Galactic dust residual in the CMB, we first measure the cross correlation between the CSFD dust map and the CMB maps. Figure \ref{fig:Fig15} shows the measured $k$NN statistic $\psi_{\textrm{CSFD-CMB}}$. The uncertainties, shown as gray shading, are estimated by cross-correlating the CSFD dust map with 1000 random CMB maps. We find no clear evidence of contamination from Galactic dust: the differences between pipelines are negligible, lying well within the scatter of the random vectors.

Estimating the CIB contamination is more challenging, because the CIB is itself our tracer of large-scale structure. However, it shares the same emission mechanism (thermal dust) as Galactic dust. The two are not strictly identical, but to first order the CIB should leak into the CMB in a manner similar to Galactic dust. Because the measured $\psi_{\textrm{CSFD-CMB}}$ shows a slight negative tendency (though not significant), a CIB residual leaking with the same sign would act to suppress rather than enhance our measured amplitude. Moreover, the CIB fluctuations are much smaller ($<2\%$, see Appendix \ref{sec:AppC} for a comparison under the mask used in the main analysis) than those of Galactic dust, so we expect the contamination to be minimal, much smaller than the CMB fluctuations themselves. \\

\subsection{Tomographic Cross Correlation: Implications for Evolving Dark Energy\label{sec:Sec5.2}}
As mentioned in Section \ref{sec:Sec1}, one of the ongoing debates about dark energy is its possible deviation from $\Lambda$CDM, such as $w$CDM ($w\neq-1$) and the time evolution of $w$ \citep[e.g.,][]{Chevallier2001, Linder2003}. In this study, we presented the relative strength of the CIB--ISW cross-correlation signal compared to the $\Lambda$CDM cosmology (MXXL simulation). Although our result is consistent with $\Lambda$CDM within $1\sigma$ (Section \ref{sec:Sec4.2}), we further use tomographic cross correlation to test for a deviation from $w=-1$. 

\begin{figure}[t]
\centering
\includegraphics[width=0.95\linewidth]{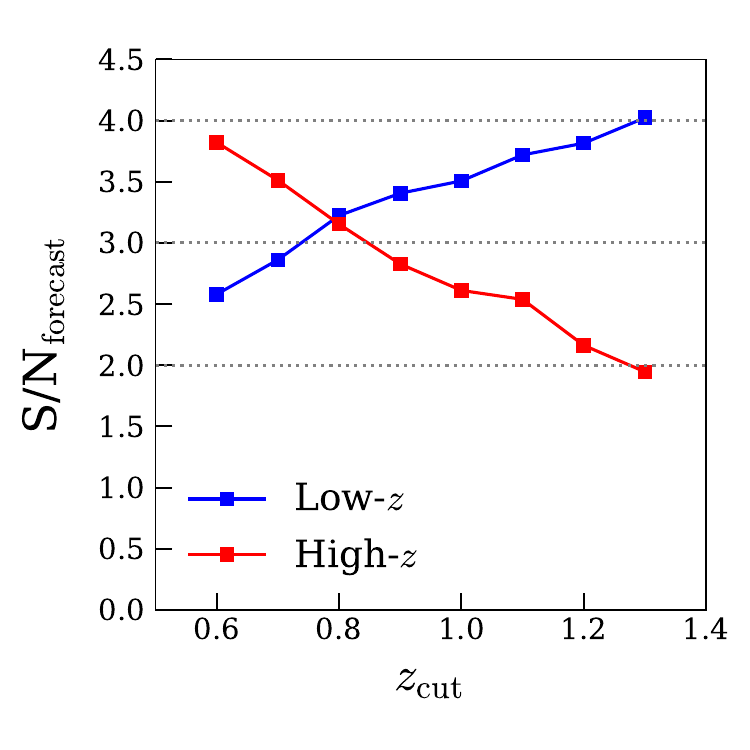}
\caption{S/N forecast of two tomographic bins as a function of redshift boundary $z_{\mathrm{cut}}$. The mock data are separated assuming a sharp boundary between two tomographic maps. The two maps reach $\mathrm{S/N}>3$ when $z_{\mathrm{cut}}\sim0.8$ is used. 
\label{fig:Fig16}}
\end{figure}

As the galaxy templates used for the CIB reconstruction span a broad redshift range (see Figure 12 of C23), the CIB can be divided tomographically. Although the template redshift distributions do not have sharp, top-hat-like boundaries, they are sufficiently narrow---especially at $z\lesssim0.5$---to allow a division into two or three tomographic bins. Because the reconstruction error increases in tomographic maps owing to the limited number of galaxy templates, we choose to construct two tomographic maps. To determine the redshift boundary $z_{\mathrm{cut}}$, we use the mock data to forecast the expected S/N. Figure \ref{fig:Fig16} shows the expected S/N values for the two redshift bins. The trend is similar to the ISW sensitivity function in Figure \ref{fig:Fig1}: the low-redshift bin requires a smaller volume to achieve the same S/N because the ISWRS effect grows with cosmic time. We adopt $z_{\mathrm{cut}}\approx0.7$, given that the reconstruction error increases with redshift.

\begin{figure*}[t]
\centering
\includegraphics[width=0.95\textwidth]{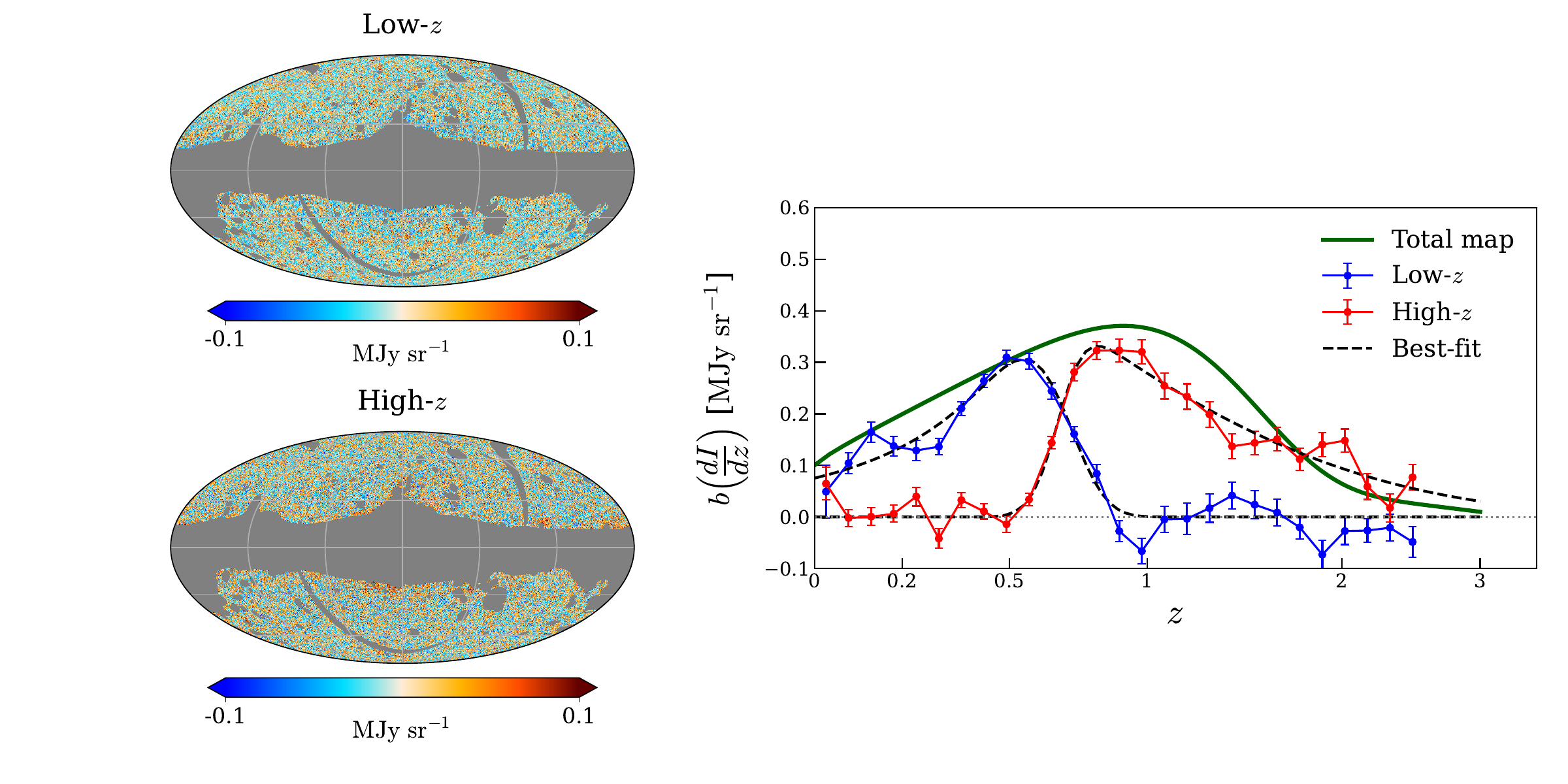}
\caption{The intensity map (left) and bias-weighted redshift distribution (right) of the two tomographic maps. The maps are divided at $z\sim0.7$ with Gaussian-like tails. The green line shows the bias-weighted redshift distribution of the full CIB map and the black dashed lines show the best fit to Equation \ref{eq:Eq16}.
\label{fig:Fig17}}
\end{figure*}

Figure \ref{fig:Fig17} shows the maps and the bias-weighted redshift distributions of two constructed tomographic bins. The two maps are divided at $z\sim0.7$ with Gaussian-like tails. The best-fit amplitude ($A_{k\mathrm{NN}}$) is measured following the same procedure presented in Sections \ref{sec:Sec2} and \ref{sec:Sec3}. The steps are summarized as follows:
\begin{enumerate}
  \renewcommand{\theenumi}{\roman{enumi}}   
  \renewcommand{\labelenumi}{(\theenumi)}   
  \item Assign infrared luminosities to MXXL lightcone halos using the W18 model.
  \item Apply a transfer function and add noise to match $C_{\ell}^{\mathrm{CIB-CIB}}$ and PDF with the observation.
  \item Estimate the covariance matrix using 200 realizations of CIB and calculate 1000 mock $\psi$.
\end{enumerate}
One difference here is the functional form used to fit the redshift distribution. We use the following empirical function for both redshift bins:
\begin{equation}
\label{eq:Eq16}
\begin{gathered}
  b\frac{dI}{dz} \propto
  \left[\left( \frac{1+\operatorname{erf}(t_L)}{2} \right)
        \left( \frac{1+\operatorname{erf}(t_R)}{2} \right)\right] \\
  t_L = \frac{z-z_L}{\sqrt{2}\sigma_L}, \quad
  t_R = \frac{z_R-z}{\sqrt{2}\sigma_R}
\end{gathered}
\end{equation}
where $\operatorname{erf}$ is the Error function and $z_L,\sigma_L, z_R, \sigma_R$ are the fitting parameters. The black dashed lines in Figure \ref{fig:Fig17} show the best-fit functions.

The best-fit amplitudes of two redshift bins are $A_{k\mathrm{NN}}^\mathrm{low\text{-}z}=1.41\pm0.47~(3.0\sigma)$ and $A_{k\mathrm{NN}}^\mathrm{high\text{-}z}=1.66\pm0.81~(2.0\sigma)$ when using the same sky coverage ($f_{\textrm{sky}}\approx0.49$). The S/N of the tomographic bins is slightly different from the forecast because of (1) reconstruction errors and (2) differences in kernel shape, including interbin redshift leakage, whereas the forecast assumed idealized top-hat kernels. Although $\sigma_{A_{k\mathrm{NN}}}$ is not small enough to clearly distinguish the tomographic results from the full-map result, both the high-$z$ and low-$z$ maps yield amplitudes higher than that of the full map and higher than the $\Lambda$CDM prediction. We note that this configuration is not unusual: the two bins are positively correlated ($r\approx0.4$) through their overlapping redshift kernels, and ${\sim}33\%$ of mock realizations with both tomographic amplitudes above unity still yield a full-map amplitude below unity. We also compare the signal with linear theory, which is identical to Section \ref{sec:Sec5.1}. The best-fit amplitudes are $A_{\mathrm{ISW}}^\mathrm{low\text{-}z}=1.06\pm0.59~(1.8\sigma)$ and $A_{\mathrm{ISW}}^\mathrm{high\text{-}z}=1.19\pm1.00~(1.2\sigma)$ when assuming Planck cosmology. Although the significance is smaller, they show a trend similar to the case of $k$NN analysis. The best-fit results are summarized in Appendix \ref{sec:AppD}.\\

\subsection{Increasing the Signal-to-noise Ratio \label{sec:Sec5.3}}
Our analysis points to several strategies for increasing the signal-to-noise ratio and directions for future improvement.
\begin{enumerate}
    \item\label{diss-redshift} \textbf{Optimal redshift distribution.} The ISW sensitivity function depicted in Figure \ref{fig:Fig1} reflects the number of galaxies and the redshift evolution of the ISW effect. It approximates the trend of ISW--density cross correlation, but the same shape of galaxy distribution does not maximize the S/N for a given total number of galaxies. \citet{Ferraro2022} have shown, using the Lagrange multiplier method combined with linear theory, that the S/N increases when galaxies are weighted toward higher redshift relative to the ISW sensitivity function. Because our analysis adopts the $k$NN statistic rather than the two-point function, the optimal kernel differs from both the ISW sensitivity function and the result of \citet{Ferraro2022}. As a check within our $k$NN framework, we reweight the mock CIB maps by a power-law kernel $W(z)\propto(1+z)^p$ normalized at $z=1$; the forecast S/N increases by $7\%$, $12\%$, and $15\%$ for $p=0.5$, $1.0$, and $1.5$. Realizing this gain in practice requires galaxy templates diverse enough to keep the CIB reconstruction reliable at high redshift, where reconstruction error grows. 
    \item \textbf{Optimal conditions for the $k$NN-CDF statistic.} The thresholds and angular scales in this study are chosen empirically. We explore $0$--$10\%$ outlier cuts on the thresholds but do not probe the full CDF range. Additional thresholds do not automatically increase S/N, as neighboring thresholds are highly correlated (Figure \ref{fig:Fig10}); a systematic optimization remains to be done. \citet{Gangopadhyay2025} noted that the number of spatial bins should be chosen to capture the full CDF shape; as long as the field model and covariance matrix are reliable, increasing the number of angular bins extracts additional information. We also conservatively remove pixels neighboring the mask to suppress bias from masked regions. Including these pixels with appropriately reduced weights would increase $f_{\textrm{sky}}$ and further improve S/N.
    \item \textbf{Synergy with future survey products.} The galaxy templates used for CIB reconstruction at high redshift are limited by the color diversity available in the WISE catalog. Multiband data from upcoming surveys such as SPHEREx \citep{Bock2026} will enable more reliable CIB reconstruction with finer tomographic resolution. SPHEREx will also make it feasible to reconstruct the CIB in multiple frequency bands over a large area; because different frequencies trace different redshift ranges, combining them increases S/N and opens a path to constraining evolving dark energy (see also point~\ref{diss-redshift}).
    \item \textbf{E-mode-assisted temperature cleaning.} The correlation between primary CMB temperature anisotropies and E-mode polarization can be exploited to partially subtract the primordial temperature component, reducing the effective temperature variance in searches for secondary signals \citep[e.g.,][]{G2012,Ballardini2019}. The achievable gain depends on the sky mask, the usable multipole range, and the polarization noise level. \citet{Goldstein2025} constructed such an E-mode-subtracted temperature map from Planck intensity and polarization data. Applying analogous cleaning to our analysis is a practical avenue for improving S/N.    \\
\end{enumerate}

\section{Conclusion \label{sec:Sec6}}
We used the $100\,\micron$ CIB and the $k$NN-CDF method to detect the ISWRS effect. We measured the $k$NN cross-correlation statistic $\psi$ using the $100\,\micron$ CIB from \citet{Chiang2023} and the Planck CMB temperature map. The measured signal was compared with the $\Lambda$CDM mock data to study the properties of dark energy. The principal results of this study are as follows.
\begin{enumerate}
    \item The mock CIB map was constructed using the halo infrared emission model presented by \citet{Wu2018}. We scaled the model up to ${\sim}60\%$ to match the bias-weighted redshift distribution from \citet{Chiang2025}. To mimic the one- and two-point statistics of the $100\,\micron$ CIB, we applied a transfer function and added white noise to the simulated maps. The PDFs of the top-hat smoothed mock CIB map are consistent with the observation. The mock ISWRS map was constructed using cubic spline interpolation of the potential fields across multiple simulation snapshots.
    \item We detected a positive cross-correlation signal between $100\,\micron$ CIB and CMB temperature. Each angular bin ($\theta=0\fdg1\text{--}0\fdg9$) deviates from the null hypothesis by $(1.8\sigma\text{--}4.1\sigma)$. Combining all 45 bins with the full covariance matrix yields $\chi_{\textrm{null}}^2\approx66.6$. The ratio of the positive correlation signal relative to the $\Lambda$CDM prediction is $A_{k\mathrm{NN}}=0.95\pm0.20$, which corresponds to $4.8\sigma$ detection of the ISWRS effect. The analysis using an angular power spectrum with the same condition (e.g., mask) yields $\lesssim4.0\sigma$ detection.
    \item We constructed two tomographic maps that have a boundary around $z_{\mathrm{cut}}\approx0.7$ to study the characteristics of dark energy. We applied the same methodology as the full-map analysis to compare the cross correlation with the $\Lambda$CDM prediction. The best-fit amplitudes were $A_{k\mathrm{NN}}^\mathrm{low\text{-}z}=1.41\pm0.47~(3.0\sigma)$ and $A_{k\mathrm{NN}}^\mathrm{high\text{-}z}=1.66\pm0.81~(2.0\sigma)$.
\end{enumerate}
With ongoing and upcoming full-sky surveys such as A-SPEC \citep{Jang2026,Kim2026,Kwon2026} and SPHEREx \citep{Bock2026}, the CIB reconstruction method combined with the $k$NN-CDF statistic will provide a powerful framework for probing dark energy and cosmic evolution.

\begin{acknowledgments}
We thank the referee for constructive comments.
D.K. is supported by the Global-LAMP Program of the National Research Foundation of Korea (NRF) grant funded by the Ministry of Education (No. RS-2023-00301976). 
D.J. was supported by NSF grants AST-2307026 and AST-2407298 at PSU, as well as by a KIAS Individual Grant, PG088301.
H.S.H. acknowledges support from the National Research Foundation of Korea (NRF) funded by the Korea government (MSIT; RS-2026-25482692) and the Global-LAMP Program funded by the Ministry of Education (RS-2023-00301976). 
Y.C. is supported by the National Science and Technology Council of Taiwan through grants NSTC 111-2112-M-001-090-MY3 and NSTC 114-2112-M-001-063-MY3, and by Academia Sinica through the Career Development Award AS-CDA-113-M01. 
This work is also supported by the Center for Advanced Computation at Korea Institute for Advanced Study. The authors would like to thank Raul Angulo, Volker Springel, and Alex Smith for kind guidance on the MXXL simulation data.
\end{acknowledgments}

\appendix

\section{Mass function and linear bias of MXXL lightcone halos \label{sec:AppA}}
\begin{figure*}[t]
\centering
\includegraphics[width=0.99\textwidth]{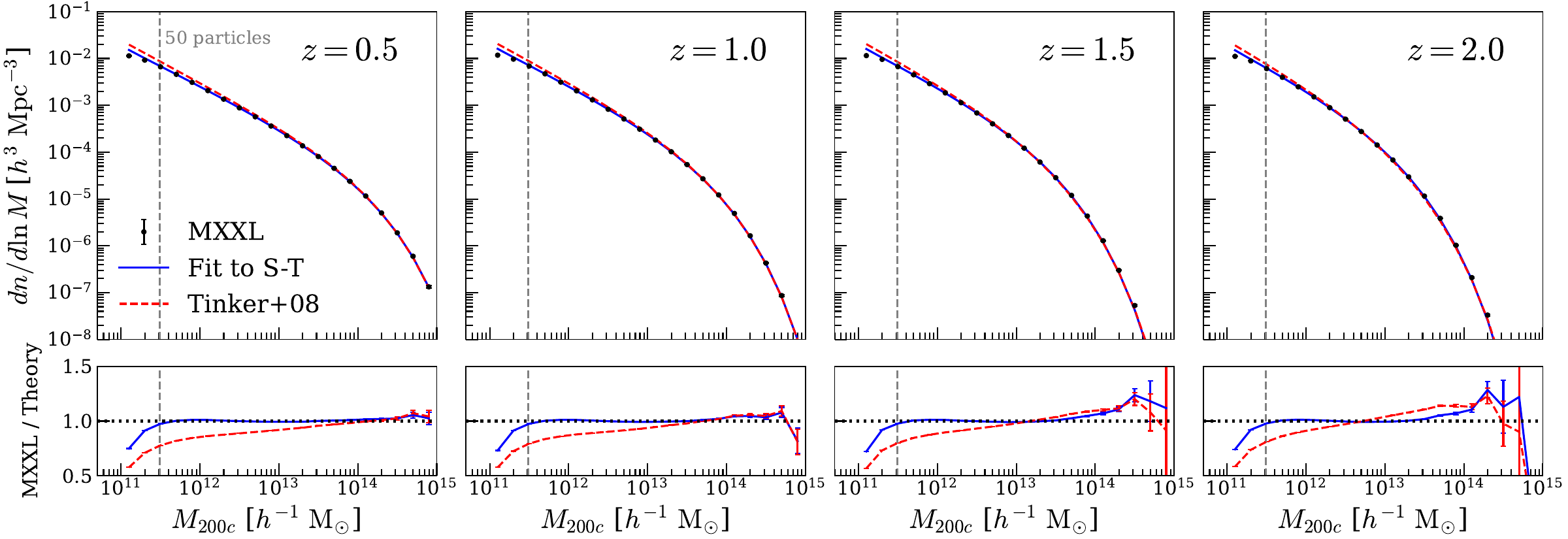}
\caption{Top: mass functions of MXXL lightcone halos (black marker) at four redshifts ($z=0.5,1.0,1.5,2.0$). We use the analytical fit (blue solid line) to the functional form of \citet{Sheth1999} (S-T in the figure). The gray dashed lines denote the mass corresponding to 50 particles. The best fit is consistent with the simulation at $10^{11.5}\,h^{-1}\,M_\sun < M_{200c} < 10^{13.5}\,h^{-1}\,M_\sun$. The analytical fitting formula from \citet{Tinker2008} (red dashed line) is shown together to demonstrate the difference.
\label{fig:App1}}
\end{figure*}

In this appendix, we present the measurements of halo mass function and bias with the MXXL lightcone halo catalog. These measurements are used in Section \ref{sec:Sec2.2.1} to generate a mock CIB map. The halos are divided into 21 spherical shells of width $\Delta z=0.1$ in the range of $z=[0.05,2.15]$. For each redshift bin, we find a best fit to the functional form of \citet[][S-T]{Sheth1999} following the same prescription of \citet{Smith2017}:
\begin{align}
    \frac{dn}{d\ln M} &= \frac{\overline{\rho}}{M} \nu f(\nu) \frac{d\sigma^{-1}}{d \ln M} \\
    \nu f(\nu) &= A \sqrt{\frac{2a}{\pi}} \left[ 1 + \frac{1}{(a\nu^2)^p} \right] \nu \exp\left( -\frac{a\nu^2}{2} \right)
\end{align}
where $M$ is the mass, $\overline{\rho}$ is the mean matter density, $\sigma^2(M)$ is the mass variance, and $\nu \equiv \frac{\delta_c}{\sigma(M)}$. We set $\delta_c=1.686$ and fit three parameters $(A,a,p)$ to the halo abundance of each redshift bin. We exclude halos containing less than $\approx60$ particles and those with $M_{200c}>6\times10^{14}\,h^{-1}\,M_{\sun}$ in the fitting process where the mass function deviates from the global trend due to the resolution or the small number of halos. Figure \ref{fig:App1} shows the MXXL halo abundance and best fit to the S-T functional form. We also show the mass function of \citet{Tinker2008} to demonstrate the difference arising from the choice of the halo finder.

\begin{figure*}[t]
\centering
\includegraphics[width=0.99\textwidth]{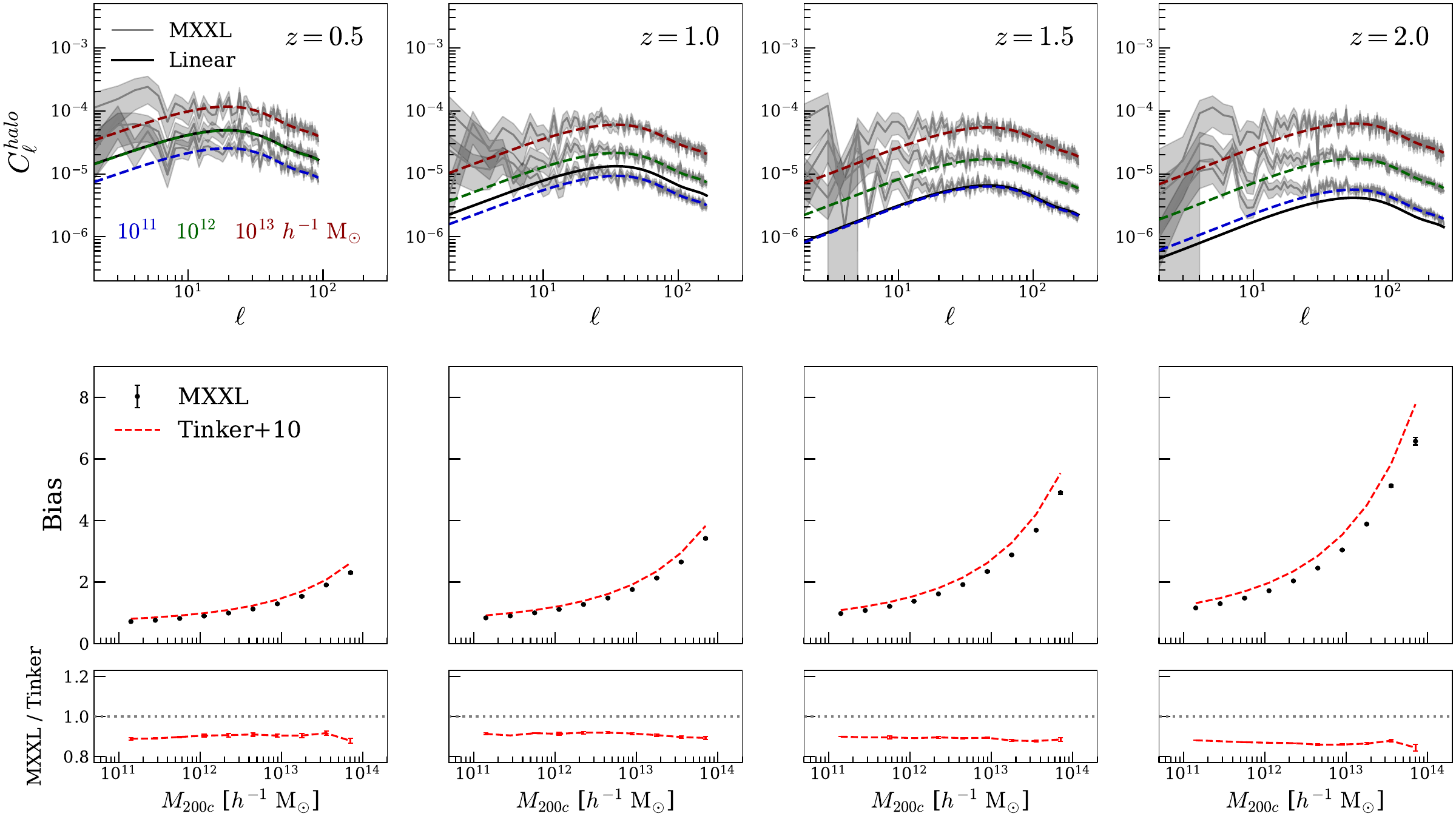}
\caption{Top: angular power spectra of MXXL lightcone halos (gray lines) for three different mass bins ($M_{200c}\sim$ $10^{11}$, $10^{12}$, $10^{13}\,h^{-1}\,M_\sun$) and four epochs ($z=$ $0.5$, $1.0$, $1.5$, $2.0$). The linear bias is measured using the best fit (dashed lines) to the power spectrum in the linear theory (black solid lines). Middle: linear halo bias as a function of mass. Bottom: ratio between measured bias and analytical fit from \citet{Tinker2010}.
\label{fig:App2}}
\end{figure*}

For the linear halo bias, we use the angular power spectrum. We construct a density contrast field using $N_{\mathrm{side}}=256$ HEALPix and calculate the power spectrum with \texttt{anafast} in HEALPix. The halos are divided into 10 mass bins and corresponding shot noise is subtracted for each measurement. To ensure that the power spectrum is in the linear regime, we apply different multipole cuts $\ell_{\max}=20\text{--}300$ when comparing the computed power spectra to the power spectrum in the linear theory obtained from the CAMB library \citep{Lewis2000}. The first row of Figure \ref{fig:App2} shows the measured MXXL halo angular power spectra (gray) and best-fit (dashed lines; $C_{\ell}^{halo}=b^2 C_{\ell}^{lin}$) at four redshifts. The best-fit $b$ as a function of mass is shown in the second row. Along with the mass function, the linear bias of MXXL halos deviates from that of \citet{Tinker2010}. For the linear halo bias, we use linear interpolation to obtain a bias at arbitrary $(M,z)$, because the quality of the analytical fit is worse than that of the mass function. \\

\section{Covariance estimation methods for cross angular power spectrum \label{sec:AppB}}
\begin{figure}[t]
\centering
\includegraphics[width=0.9\linewidth]{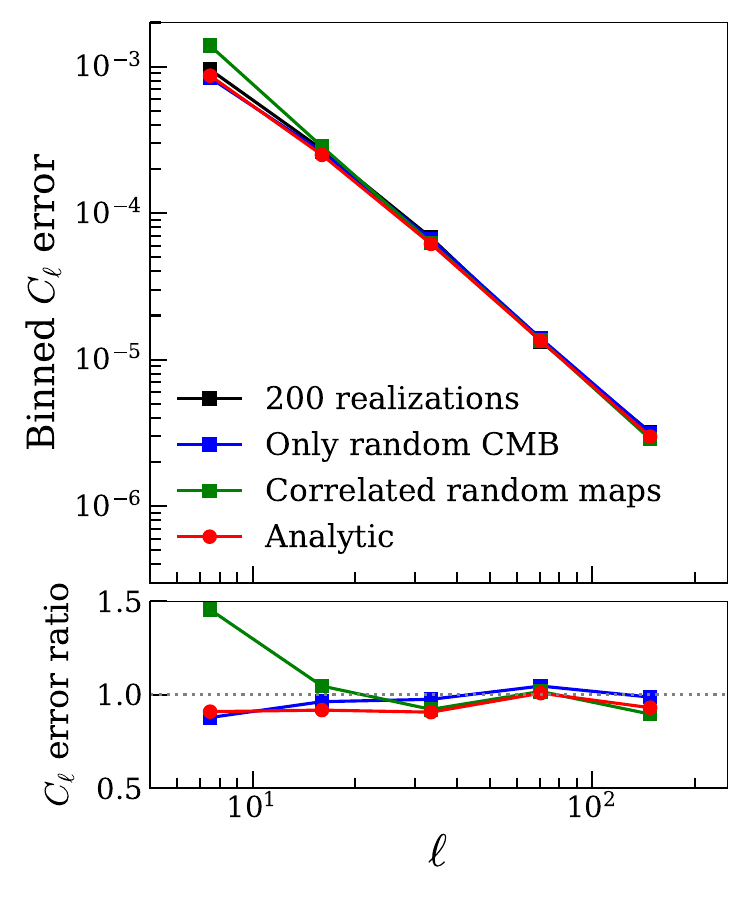}
\caption{Top: binned error of cross angular power spectrum $C_{\ell}^{\mathrm{CMB-CIB}}$ depending on the covariance estimation method. The black line shows the original method used in the main text, namely 200 realizations of CIB cross-correlated with random CMB. The blue line denotes the case where only random CMB is considered. The green line is the estimation using 1000 correlated random maps of CIB and (ISW+CMB). The red line shows the analytical covariance. Bottom: ratio compared to the original method. The diagonal terms of the covariance matrices vary by up to ${\sim}40\%$, but the variation in the final S/N is less than $7\%$ as shown in Table \ref{tab:tab3}. 
\label{fig:App3}}
\end{figure}

\begin{deluxetable}{lccc}
\tablecaption{Dependence of $A_{\mathrm{ISW}}$ and S/N on the covariance estimation method.\label{tab:tab3}}
\tablecolumns{3}
\tablewidth{0pt}
\tablehead{
\multicolumn{1}{l}{Method} & \colhead{$A_{\mathrm{ISW}}$} & \colhead{S/N} & \colhead{S/N Ratio}
}
\startdata
200 realizations of CIB  & $0.97\pm0.25$ & $3.90$ & \nodata \\
Only random CMB          & $0.90\pm0.23$ & $3.99$ & 1.02 \\
Correlated random maps   & $1.08\pm0.27$ & $3.97$ & 1.02 \\
Analytic                 & $0.96\pm0.23$ & $4.17$ & 1.07 \\
\enddata
\tablecomments{$A_{\mathrm{ISW}}$: best-fit amplitude; S/N: detection significance.}
\end{deluxetable} 
In this appendix, we repeat our angular power spectrum analysis presented in Section \ref{sec:Sec5.1} with different covariance estimation methods. In the main text, the covariance matrix is estimated by cross-correlating the 200 realizations of the CIB and random CMB (both $C_{\ell}$ and $\psi$). Here, we use three different methods: (1) CIB map and 200 random CMB (no reconstruction error), (2) correlated random maps of CIB and (ISW+CMB) using linear theory, and (3) analytical covariance. The correlated random maps are generated with \texttt{synfast} in HEALPix using Equations (\ref{eq:Eq2}) and (\ref{eq:Eq15}). The analytical covariance matrix is as follows:
\begin{equation}
    \Delta^2 C_{\ell}^{TI}=\frac{C_{\ell}^{TT}\,C_{\ell}^{II}+(C_{\ell}^{TI})^2}{(2\ell+1)f_{\textrm{sky}}}
\end{equation}
where $T$ and $I$ in the superscript denote CMB and CIB respectively. The results are summarized in Figure \ref{fig:App3} and Table \ref{tab:tab3}. All methods yield S/N values consistent with the fiducial measurement to within $7\%$. The fiducial covariance adopted in the main text (200 realizations of CIB) gives the most conservative (lowest) S/N among the methods considered. Neglecting the CIB reconstruction error (only random CMB) underestimates the uncertainty and increases the S/N by ${\sim}2\%$, while the analytical Gaussian covariance gives the largest S/N (${\sim}7\%$ higher). The overall agreement confirms that our detection significance is robust to the choice of covariance estimation method.\\

\section{Comparison of the CSFD and CIB fluctuations \label{sec:AppC}}
\begin{figure}[t]
\centering
\includegraphics[width=0.9\linewidth]{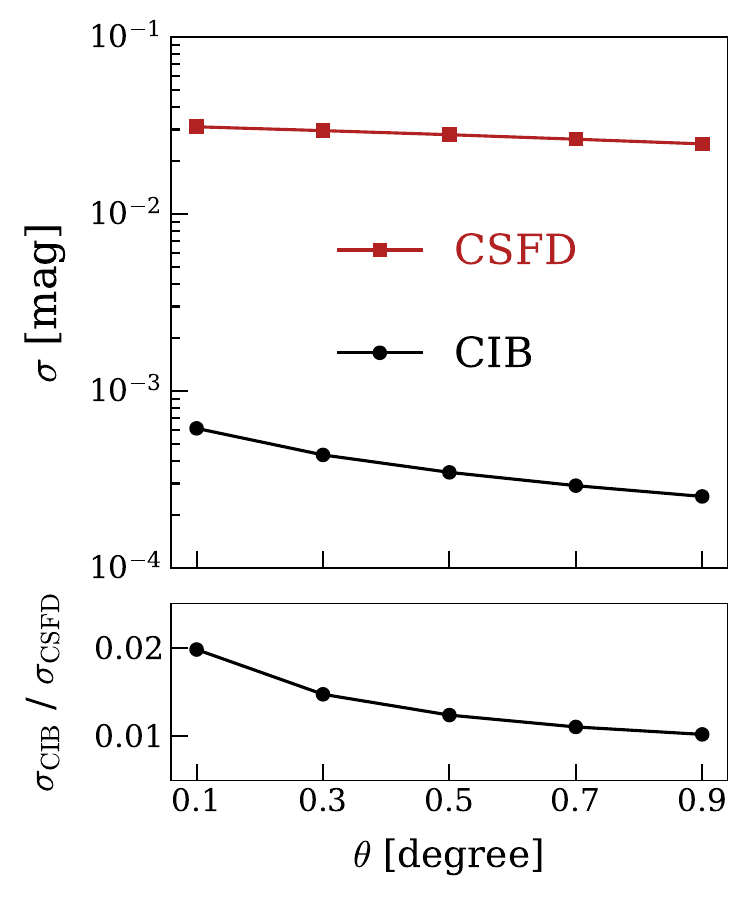}
\caption{Top: standard deviation of top-hat smoothed CSFD and CIB maps as a function of angular scale $\theta$. Bottom: the ratio of the CIB to the CSFD fluctuations.
\label{fig:App4}}
\end{figure}

In this appendix, we compare the fluctuations between the CSFD dust map and the C23 CIB map within the sky area used for the main analysis ($f_{\mathrm{sky}}\approx0.49$). The two maps are smoothed with top-hat filters of radii $\theta\in[0.1\arcdeg,0.9\arcdeg]$ as in the main $k$NN analysis. The upper panel of Figure \ref{fig:App4} shows the standard deviation of each map, $\sigma$, as a function of angular scale $\theta$. Although we use only the reliable sky area far from the Galactic disk, the CIB fluctuations are $<2\%$ of the Galactic dust level (bottom panel).

\section{Best-fit amplitudes for tomographic maps \label{sec:AppD}}
\begin{figure*}[t]
\centering
\includegraphics[width=0.99\textwidth]{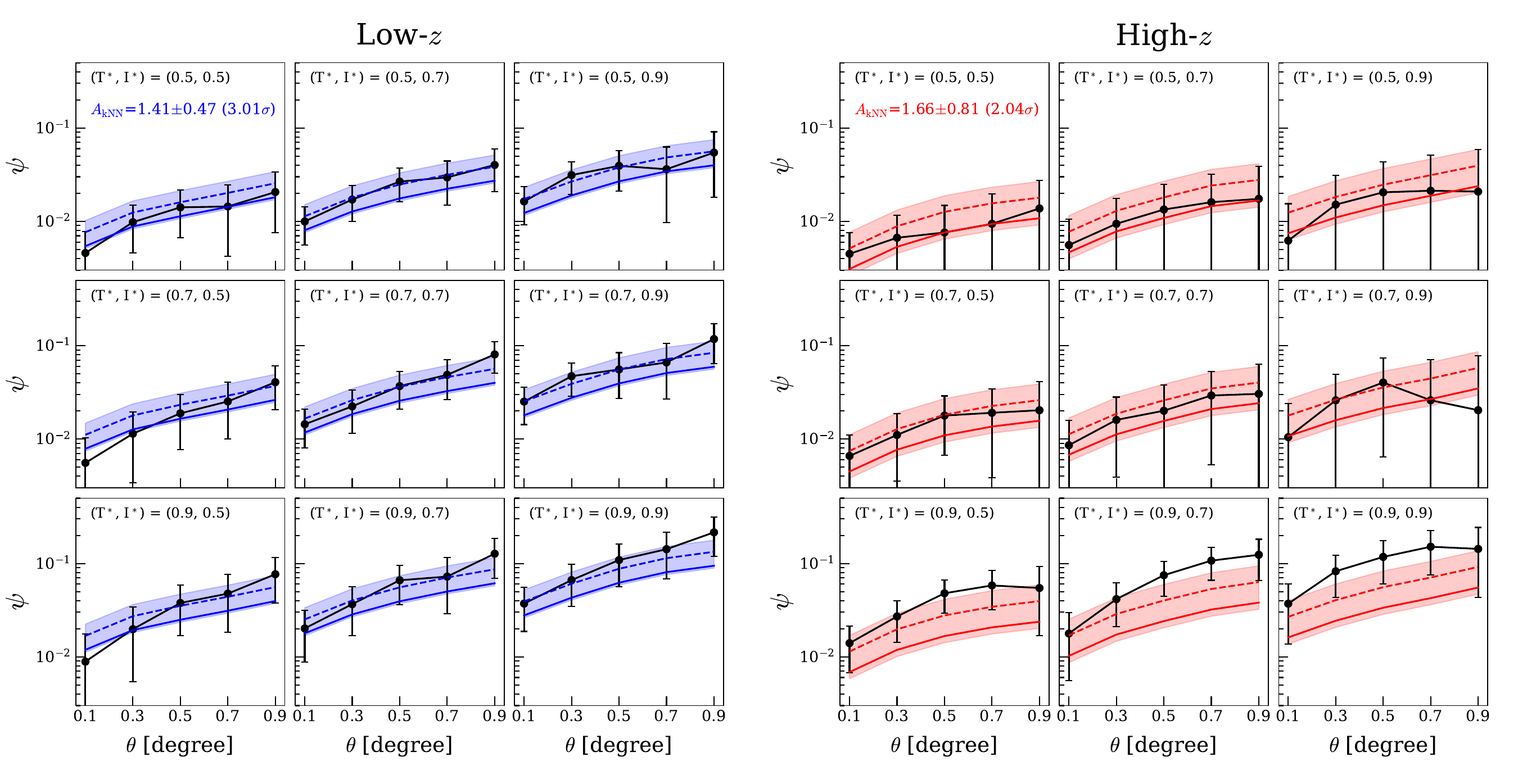}
\caption{$k$NN statistic $\psi$ measured with two tomographic maps (left: low-$z$, right: high-$z$). Black markers denote the measurements from the observational data. Solid lines are the predictions from the MXXL mock data. Dashed lines and shading show the best-fit amplitudes and corresponding $1\sigma$ uncertainty.
\label{fig:App5}}
\end{figure*}

\begin{figure*}[t]
\centering
\includegraphics[width=0.88\textwidth]{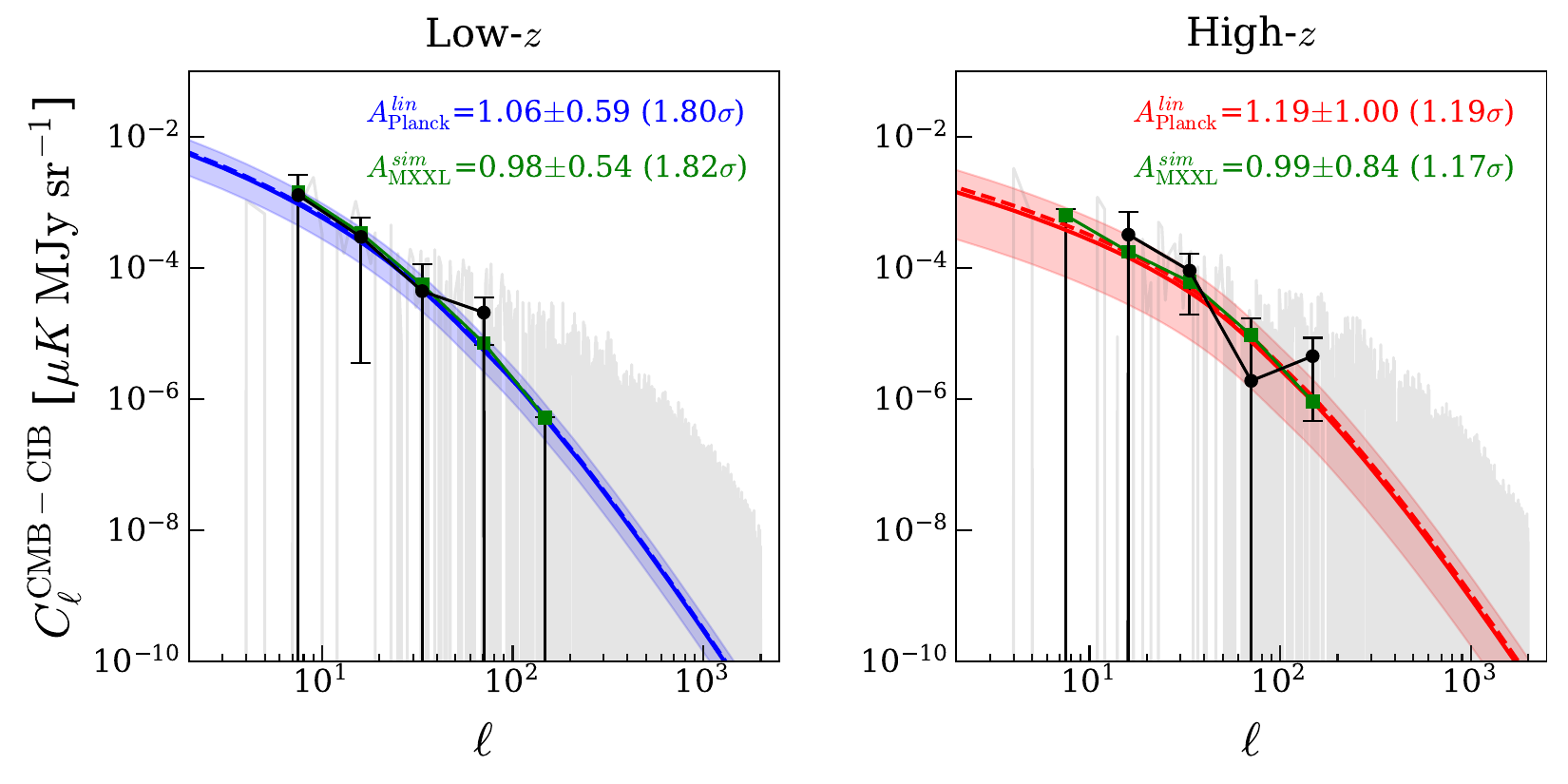}
\caption{Same as Figure \ref{fig:App5}, but for cross angular power spectrum $C_{\ell}^{\mathrm{CMB-CIB}}$.
\label{fig:App6}}
\end{figure*}

In this appendix, we demonstrate the full fitting result of the tomographic maps for both $k$NN statistic and angular power spectrum. We use approximately the same sky fraction $f_{\textrm{sky}}\approx0.49$ and outlier cuts $(0,0.02)$ for the $k$NN statistic as in the full-map analysis. Figure \ref{fig:App5} shows the result using the $k$NN statistic. The left (right) panel is similar to Figure \ref{fig:Fig13}, but for the low-$z$ (high-$z$) map. The black data points are measurements from the observation and the solid lines are from MXXL. The dashed lines and shading show the best-fit amplitude and corresponding $1\sigma$ uncertainty. Figure \ref{fig:App6} shows a similar measurement, but with cross angular power spectrum $C_{\ell}^{\mathrm{CMB-CIB}}$.

\bibliography{CIB_ISW_kNN}{}
\bibliographystyle{aasjournalv7}



\end{document}